\documentclass[fleqn,usenatbib]{mnras}
\usepackage[T1]{fontenc}

\usepackage{newtxtext,newtxmath}

\usepackage{ae,aecompl}
\usepackage{graphicx}	
\usepackage{amsmath}	
\usepackage{bm}	
\usepackage{hyperref}
\newcommand{\hMsun}{ h^{-1}{\rm M_{ \odot}}}
\newcommand{\hMpc}{ h^{-1}{\rm Mpc}}

\newcommand{\ihMpcC}{ h^{3}{\rm Mpc}^{-3}}

\newcommand{\vpeak}{V_{\rm peak}}
\newcommand{\mpeak}{M_{\rm peak}}
\newcommand{\minfall}{M_{\rm infall}}
\newcommand{\Fs}{f_{\rm s}}
\newcommand{\sigmaLogM}{\sigma_{\rm logM}}
\newcommand{\tmerger}{t_{\rm merger}}

\title[A flexible SHAM model]{A flexible subhalo abundance matching model for galaxy clustering in redshift space}

\author[S. Contreras et al.]{
S. Contreras,$^{1}$\thanks{E-mail: sergio.contreras@dipc.org}
R. E. Angulo,$^{1,2}$
 \& M. Zennaro$^{1}$.
\\
$^{1}$Donostia International Physics Center (DIPC), Manuel Lardizabal Ibilbidea, 4, 20018 Donostia, Gipuzkoa, Spain.\\
$^{2}$IKERBASQUE, Basque Foundation for Science, 48013, Bilbao, Spain.
}

\date{Accepted XXX. Received YYY; in original form ZZZ}

\pubyear{2020}

\begin{document}
\label{firstpage}
\pagerange{\pageref{firstpage}--\pageref{lastpage}}
\maketitle 

\begin{abstract}
We develop an extension of subhalo abundance matching (SHAM) capable of accurately reproducing the real and redshift-space clustering of galaxies in a state-of-the-art hydrodynamical simulation. Our method uses a
low-resolution gravity-only simulation and it includes orphan and tidal disruption prescriptions for satellite galaxies, and a flexible amount of galaxy assembly bias. Furthermore, it includes recipes for star formation rate (SFR) based on the dark matter accretion rate. We test the accuracy of our model against catalogues of stellar-mass- and SFR-selected galaxies in the TNG300 hydrodynamic simulation. By fitting a small number of free parameters, our extended SHAM reproduces the projected correlation function and redshift-space multipoles for number densities $10^{-3} - 10^{-2}\,\ihMpcC$, at $z=1$ and $z=0$, and for scales $r \in [0.3 - 20]\hMpc$. Simultaneously, the SHAM results also retrieve the correct halo occupation distribution, the level of galaxy assembly bias, and higher-order statistics present in the TNG300 galaxy catalogues. As an application, we show that our model simultaneously fits the projected correlation function of the SDSS in 3 disjoint stellar mass bins, with an accuracy similar to that of TNG300 galaxies. This SHAM extension can be used to get accurate clustering prediction even when using low and moderate-resolution simulations.
\end{abstract}


\section{Introduction}
\label{sec:Introduction}

With a new generation of surveys soon to start collecting data, a new generation of theoretical predictions for galaxy clustering is needed. These ``mock'' galaxies are essential to prepare for these observations, estimate sources of uncertainty, and exploit them scientifically. Due to the large volume and statistical power of upcoming surveys, this new generation of mocks will need to be accurate and computationally efficient.

In the $\Lambda$CDM paradigm, galaxies form from the gas that is captured in the gravitational potential well of dark matter haloes \citep{WhiteRees:1978}. As this gas cools, it falls to the centre of haloes and settles into a disk, to then cool and fragment into stars. While the evolution of a halo is somewhat simple, growing in mass through accretion and mergers determined mostly by gravitational interactions, the evolution of galaxies is much more complex, with a large number of additional astrophysical processes involved. Hydrodynamic simulations are probably the most realistic method to model these processes, however they can take up to hundreds of millions of CPU hours to execute as they require high spatial and mass resolution. This implies that most of these simulations have only reached volumes of the order of 100 $\hMpc$ a side (eg. EAGLE, \citealt{Schaye:2015}, Illustris,\citealt{Illustrisa}, HorizonAGN, \citealt{HorizonAGN}, Simba, \citealt{SIMBA}, simulations). 

Other techniques which simplify the modelling of some astrophysical processes, like semi-analytical models of galaxy formation \citep[SAMs,][]{Henriques:2015,Stevens:2016,Lacey:2016,Croton:2016,Lagos:2018} or semi-empirical models such as EMERGE \citep{Moster:2018} or The Universe Machine \citep{Behroozi:2019} can be run in post-processing over $\rm Gpc$ scale gravity-only simulations in a hundreds of CPU hours (only run, not calibrate them). While much faster than hydrodynamic simulations, this is still too slow when thousands or even millions of mocks are needed to, e.g. scan high dimensional parameter spaces or compute covariance matrices. An additional limitation is that these approaches require underlying simulations with considerably high force and mass resolution.

Although the evolution of galaxies and haloes is different, we expect a strong correlation between their properties, which motivates ``empirical models''  as an alternative to create fast and computationally efficient mocks. These models ``paint'' galaxies in haloes or subhaloes using theoretically-motivated relations between the galaxies and their host (sub)haloes. These methods can be run in a few minutes over large cosmological volumes, and have proved to successfully reproduce several galaxy observables, such as the galaxy two-point correlation function. One example of an empirical models is the SubHalo Abundance Matching technique \citep[SHAM,][]{Conroy:2006,Reddick:2013,ChavesMontero:2016,Lehmann:2017,Dragomir:2018}. This technique is based on the idea that most massive subhaloes should host the most massive/luminous galaxies. The algorithm matches, with some scatter, the selected subhalo property to the expected stellar mass/luminosity function. While simple, the model is capable of reproducing the galaxy clustering in complex galaxy formation models, such as hydrodynamic simulations (e.g. \citealt{ChavesMontero:2016}).

The main limitation of SHAM is that it assumes a direct relation between the fate of a dark matter satellite and the putative galaxy it might host. This assumption does not hold in detail, as the presence of a galaxy would modify e.g. dynamical friction timescales and/or the resilience of a subhalo against tidal disruption. Furthermore, in relatively low resolution simulations, a subhalo might be stripped below the resolution limit of the simulation, but the hosted galaxy is still expected to survive as a satellite. Another shortcoming of SHAM is that it is limited to properties expected to correlate strongly with dark matter mass, so it cannot make predictions for star-formation rates, which would be relevant for upcoming galaxy surveys targeting emission line galaxies.


In this paper, we develop an extension to the standard subhalo abundance matching technique that, with only 3 free parameters (5 if also including assembly bias) and a low-resolution dark matter only (DMO) simulation, can reproduce the clustering of stellar mass-selected galaxies in real and redshift space with high precision. This extension includes the use of {\it orphan satellite} subhaloes/galaxies tracking structures below the resolution level of the simulation; a {\it disruption mechanism} to account for satellite galaxies that should have been disrupted to form part of the intracluster medium, or in which their stellar mass has decreased so much that they should no longer be part of the galaxy selection; and a {\it flexible amount of galaxy assembly bias} to account for any possible difference between correlation with large-scale environment in SHAM and our target sample.

Along with these improvements, we develop a method capable of predicting the clustering of SFR-selected galaxies. This method assumes that the SFR of each galaxy is proportional to the dark matter accretion rate of its host halo, and that is modulated by an efficiency that depends on the mass of its host halo. This approach is similar to that employed by the semi-empirical model {\tt EMERGE} but adapted to not use the full merger tree on its computation but instead properties readily available in SHAM.

To test the performance of our SHAM extension, we use the Illustris TNG300 magneto-hydrodynamic simulation \citep{TNGa}, to our knowledge, the largest publicly cosmological hydrodynamic simulation available today ($L=205\ \hMpc$). For the stellar mass-selected mocks, we find that our method is capable of reproducing almost perfectly the real and redshift-space clustering, as well as the galaxy assembly bias level, halo occupation distribution (HOD), and higher-order statistics as quantified by the k-nearest neighbour cumulative distribution functions (kNN-CDF). Similarly, for the SFR-selected sample, we find that we are also able to reproduce the same galaxy statistics, albeit with a somewhat less accuracy, particularly for the level of assembly bias. The differences in the galaxy clustering between the star-forming mock and the TNG300 are still much smaller than those found among different galaxy formation models \citep{C13}. We expect this new generation of mocks to be useful for the development and understanding of current and future galaxy surveys.

The outline of this work is as follows: In section~\ref{sec:models} we present the simulations used in this work and introduce the SHAM technique. Our modifications to SHAM regarding the treatment of satellite galaxies are discussed in section~\ref{sec:sat} and those regarding galaxy assembly bias in section~\ref{sec:ab}. In section~\ref{sec:sfr} we present our star-formation-rate modelling. We compare the 2-point clustering of our mock galaxies with that of the TNG300 in section~\ref{sec:clustering} and with observations in section~\ref{sec:obs}. To finalise, we present our conclusions in section~\ref{sec:Conclusions}.

Unless otherwise stated, the units in this paper are $\hMsun$ for masses, $\hMpc$ for distances, $\rm km/s$ for the velocities and $\rm M_{\odot}/yr$ for star formation rates.

\section{Simulations and the standard abundance matching model}
\label{sec:models}

\subsection{The TNG300 }
\label{sec:TNG300}

To test the accuracy of our mocks, we will compare their clustering against those of galaxies in the TNG300 magneto-hydrodynamic simulation, the largest of ``The Next Generation'' Illustris Simulations suite, and, to our knowledge, the largest volume high-resolution hydrodynamic simulation publicly available \citep{TNGa, TNGb, TNGc, TNGd, TNGe}. 

The TNG300 simulation follows a periodic box of 205 $\hMpc$ ($\sim 300$ Mpc) a side. It used $2500^3$ dark matter particles and gas cells, implying a baryonic mass resolution of $7.44\times10^6\,\hMsun$ and of $3.98\times 10^7\,\hMsun$ for dark matter. The simulations were carried out using the {\tt AREPO} code \citep{AREPO} adopting cosmological parameters consistent with recent analyses \citep{Planck2015}\footnote{$\Omega_{\rm dm}$ = 0.3089, $\Omega_{\rm b}$ = 0.0486, $\sigma_8$ = 0.8159, $n_s$ = 0.9667 and $h$ = 0.6774}. While main results of this paper are only shown at $\rm z=0$, we tested that they are also valid at $\rm z=1$ (Appendix B). The outputs of the simulation are publicly available at the TNG project webpage\footnote{\url{https://www.tng-project.org/}}. 

We built galaxies catalogues by selecting the most massive or star-forming galaxies such that their abundance is equal to a number density of ${\rm n=0.01,\ 0.0316\ \&\ 0.001}$ $\ihMpcC$, equivalent to a minimum value of the stellar mass of $8.26\times 10^{9}, 2.93\times 10^{10}$ \& $ 6.46\times 10^{10}$ $\hMsun$ and a minimum SFR of 0.468, 1.493 $ \& $ 3.033 $\rm M_{\odot}/yr$, respectively. We defined the stellar mass of a galaxy as the sum of the mass of all particles within the stellar half mass radius. We also test defining the stellar mass of the galaxies as the sum of the masses of all particles of a subhalo, finding similar results. We define the SFR of a galaxy as the sum of the individual star formation rate of all gas cells in its subhalo.

We note that although the TNG300 simulation has shown to agree with many observables \citep{TNGb}, it is not the aim of this work to create mocks that only resemble this simulation, but to create a flexible model that can describe a broader range of feasible galaxy formation physics and their correlation with cosmological parameters. We chose an hydrodynamic simulation as a benchmark since it models baryons and dark matter jointly, unlike other approaches that can reproduce galaxy clustering (such as HODs, semi-empirical models or semi-analytical models). This adds an extra challenge for mocks created in DMO simulations, allowing us to perform a stringent test of the performance of our SHAM galaxies.

In addition to the TNG300, we also use the TNG300-1-Dark, TNG300-2-Dark \& TNG300-3-Dark simulations. These are DMO simulations employing the same initial white noise field as the TNG300 run. The TN300-1-Dark has the same number of dark matter particles as the TNG300 ($2500^3$) whereas the TNG300-2-Dark and TNG300-3-Dark have factors of $2^3$ and $4^3$ fewer particles (i.e. $1250^3$ and $625^3$ particles, respectively). For some comparisons, we will use the cross-match of TNG galaxies and subhaloes in the TNG300-1-Dark, available in the Illustris TNG website.

\begin{figure}
\includegraphics[width=0.45\textwidth]{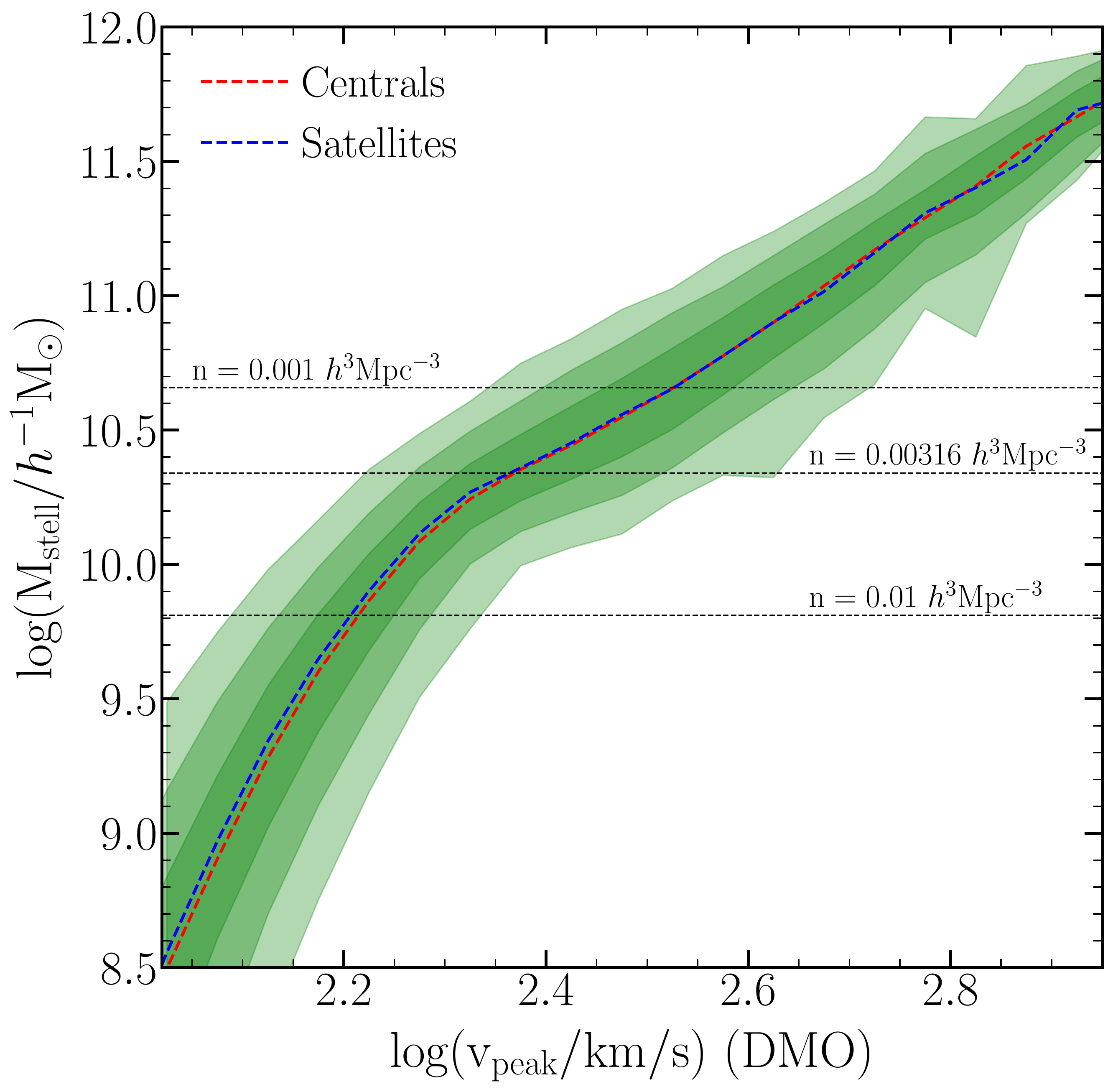}
\caption{
The stellar mass of the galaxies of the TNG300 simulation as a function of  $\vpeak$ of the subhaloes of the matched substructures in the TNG300-1-Dark simulation. The shaded regions represent 1, 2 \& 3 $\sigma$ of the distribution, and the dashed red and blue lines show the median of the central and satellite galaxies, respectively. As a reference, the three horizontal dotted lines mark the three number densities used in this work, as labeled.}
\label{Fig:fig01}
\end{figure}

\begin{figure*}
\includegraphics[width=1.00\textwidth]{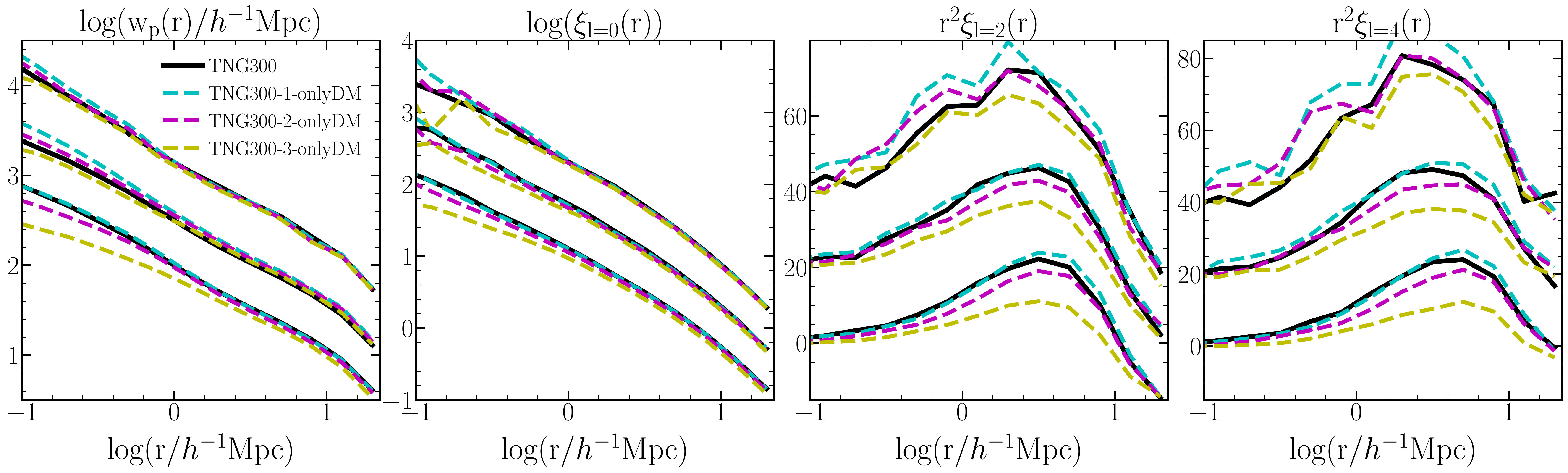}
\caption{
The projected correlation function ($\rm w_p(r_p)$, rightmost panel), the monopole ($\rm \xi_{\ell=0}(r)$, second panel), quadrupole ($\rm \xi_{\ell=2}(r)$, third panel) and hexadecapole ($\rm \xi_{\ell=4}(r)$, leftmost panel) of the galaxies of the TNG300 simulation (black solid line) and the subhaloes of the TNG300-1-Dark, TNG300-2-Dark \& TNG300-3-Dark dark matter only simulation (cyan, magenta and yellow dashed lines, respectively). The galaxies/subhaloes samples have a number density of n= 0.01, 0.00316 \& 0.001 $h^3{\rm Mpc^{-3}}$  selected by stellar mass (for the TNG300) and $\vpeak$ (for the dark matter only simulations). For visualisation propose, the lines for the two lowest number densities have been displaced in the y-axis. The lines in $\rm w_p(r_p)$ and $\rm \xi_{\ell=0}(r)$ have been displaced in 0.7 and 1.4, and 20 and 40 for the lines in $\rm \xi_{\ell=2}(r)$ and $\rm \xi_{\ell=4}(r)$, with  the lowest number density is in the top.}
\label{Fig:fig02}
\end{figure*}


\subsection{TNG300-3 mimic}
\label{sec:mimic}

The main dark matter simulation we will employ is analogous to the TNG300-3-Dark (same volume and initial conditions as the TNG300, but $4^3$ fewer particles), but with additional properties computed for its subhaloes as we describe below. We refer to this simulation as the TNG300-3-mimic, and highlight that by using a simulation with the same initial condition as the TNG300 we can perform more accurate comparisons as it reduces any difference between the model caused by cosmic variance.

The simulation was carried out with an updated version of {\tt L-Gadget3} \citep{Angulo:2012}, a lean version of {\tt GADGET} \citep{Springel:2005} used to run the Millennium XXL simulation and the Bacco Simulations \citep{Angulo:2020}. This version of the code allows an on-the-fly identification of haloes and subhaloes using a Friend-of-Friend algorithm \citep[{\tt FOF}][]{Davis:1985} and an extended version of {\tt SUBFIND} \citep{Springel:2001}. Our updated version of {\tt SUBFIND} can better identify substructures by considering the information of its past history, while also measuring properties that are non-local in time such as the peak halo mass ($\mpeak$), peak maximum circular velocity ($\rm \vpeak$), infall subhalo mass ($\minfall$), and mass accretion rate among others. 

We note that we employ the same numerical accuracy parameters as the Bacco simulations \citep{Angulo:2020}. And as a validation test, we have compared the $z=0$ mass power spectrum of our simulation to that measured in the TNG300-3-Dark finding differences below 0.8\% at all scales.

\subsection{Subhalo abundance matching}
\label{sec:SHAM}

In its most basic incarnation, SHAM assumes a one-to-one mapping between the mass of subhaloes and the stellar mass or luminosity of hypothetically hosted galaxies. More recent implementations of SHAM include scatter to this mapping and employ other subhalo properties, such as $M_{\rm infall}$, $M_{\rm peak}$, $V_{\rm max}$ or $V_{\rm peak}$, but still it is assumed that the position and fate of galaxies is determined by the evolution of dark matter subhaloes. This, however, cannot be true in detail since subhalos might be artificially disrupted in $N$-body simulations (because of numerical artifacts and/or limited mass resolution). More generally, the presence of a galaxy can modify the subhalo density profile, affecting its orbit, tidal stripping, and eventual disruption. This puts a limit to the accuracy of SHAM, which we will illustrate next.    

Fig.~\ref{Fig:fig01} compares the stellar mass of TNG300 galaxies and the value of $\vpeak$, the peak value of the maximum circular velocity of a halo throughout its life, of the corresponding cross-matched subhalo in the TNG300-1-Dark. Note that \cite{ChavesMontero:2016} showed that in the EAGLE simulation \citep{Schaye:2015} $\vpeak$ is one of the properties that best predicts the value of stellar mass. Indeed, we can see that there is a strong correlation between these two properties in TNG300, with a mean relation being almost identical for centrals and satellite galaxies, validating a key SHAM assumption. \footnote{Note that, although not shown here, we found a similar scatter between $\mpeak$ and stellar mass but with a significantly larger difference between centrals as satellite galaxies, which further motivates the use of $\vpeak$ instead of $\mpeak$ \citep[see also the discussion in][]{Campbell:2018}.} 

\cite{ChavesMontero:2016} also found that SHAM using $\vpeak$ was capable of reproducing the clustering of stellar-mass selected galaxies in EAGLE with very high accuracy. However, these authors used a very high resolution dark-matter simulation (which would imply artificial disruption of subhalos was negligible), and the accuracy on small scales varied depending on the assumptions regarding star formation in satellites and their tidal disruption. 

We illustrate the impact of mass resolution for SHAM in Fig.~\ref{Fig:fig02}. We show the projected correlation function, $\rm w_p(r_p)$, and the monopole, quadrupole, and hexadecapole of the redshift space correlation function ($\rm \xi_{\ell=0}(r)$, $\rm \xi_{\ell=2}(r)$, $\rm \xi_{\ell=4}(r)$, respectively) for stellar-mass selected galaxies at $z=0$. Solid lines show the results for TNG galaxies, whereas coloured lines show the SHAM predictions as computed using the TNG300-1-Dark, TNG300-2-Dark \& TNG300-3-Dark simulations. Note we display results for three different number densities.

We see that SHAM in the TNG300-1-Dark agrees remarkably well with the TNG measurements for the densest galaxy sample, but it overpredicts the clustering for lower number densities. Conversely, SHAM in the lowest resolution simulation, TNG300-3-Dark, agrees very well with the sparsest sample but instead heavily underpredicts the clustering of the densest sample. Note that these problems are relatively confined to small scales in real space, but can affect even large scales in the case of the quadrupole and hexadecapole.

All this is simply a consequence that in standard SHAM, the abundance of satellites depends on the numerical resolution of the parent dark matter simulation which might result in an excess or lack of satellite galaxies. For instance, galaxies in the TNG300 can be affected by stripping of the stellar mass, while subhaloes never decrease their $\vpeak$, producing an overabundance of satellite galaxies \citep{Smith:2016}. Overall, this highlights an important limitation of SHAM and motivates our developent of an extension to SHAM that addresses these problems.

\label{sec:SHAM}
\begin{figure*}
\includegraphics[width=1.\textwidth]{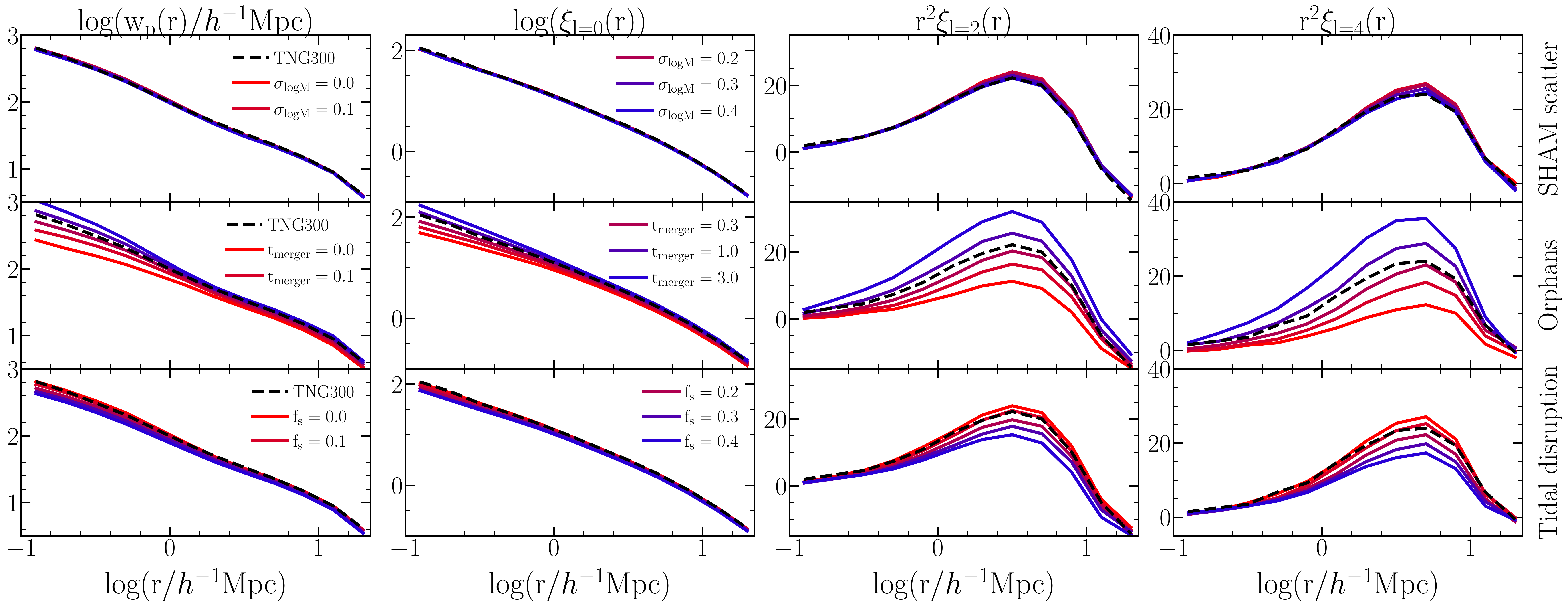}
\caption{ The impact of the free parameters in our extended SHAM for stellar-mass selected galaxies with a number density of n = 0.01 $h^3{\rm Mpc^{-3}}$ at $z=0$. The first column displays the projected correlation function, $\rm w_p(r_p)$; whereas the second, third and fourth columns respectively display the monopole $\rm \xi_{l=0}(r)$, quadrupole $\rm \xi_{\ell=2}(r)$, and hexadecapole $\rm \xi_{\ell=4}(r)$ of the redshift space correlation function. In each row we systematically vary a different SHAM parameter, $\sigmaLogM$, $\tmerger$, and $\Fs$, while keeping all the other parameters fixed. Red to blue colours indicate changes from low to high parameter values, as indicated by the legend.}
\label{Fig:fig4}
\end{figure*}

\begin{figure*}
\includegraphics[width=1.\textwidth]{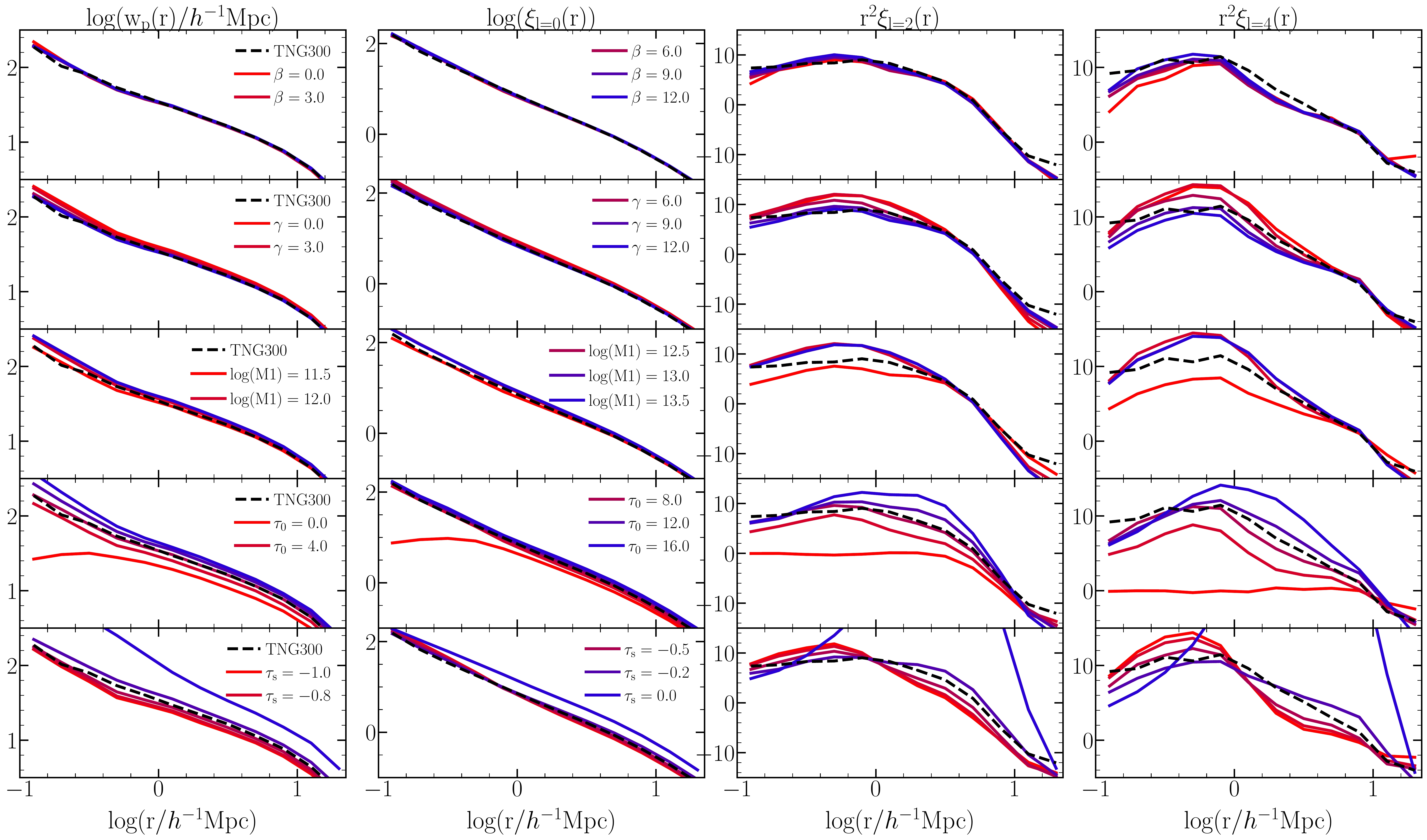}
\caption{Similar to Fig.~\ref{Fig:fig4} but for the SFR parameters: the slopes in the star formation efficiency ($\beta$ \& $\gamma$), the halo mass of the peak of the star formation efficiency ($M_1$), the parameters that control de abundance of satellite galaxies $\tau_0$ and its dependence with it host halo mass $\tau_S$.}
\label{Fig:fig4p5}
\end{figure*}

We use the subhalo abundance matching technique (SHAM) with $\vpeak$ as a starting point of our mocks. Firstly, we match the value of $\vpeak$ of the dark matter only simulation to the stellar mass function of the TNG300 simulation. Since most of our analysis is done using cuts in number density, the results are mostly independent of the stellar mass function used. We then i) add a free parameter controlling the scatter ($\sigmaLogM$) between the stellar mass of galaxies and their $\vpeak$, ii) explicitly model the level of assembly bias (following the procedure presented in \citealt{C20b}), and iii) regulate the disruption of satellites and the star formation rate with recipes inspired by semi-analytic galaxy formation models and by the semi-empirical model {\tt EMERGE} \citep{Moster:2018}. In the next sections we will discuss our prescriptions and show the impact that the free parameters of our model have on the predicted redshift space clustering of galaxies.

\section{The satellite treatment in SHAM}
\label{sec:sat}

In this section, we introduced two of SHAM modifications related to the treatment of satellite galaxies. In section~\ref{sec:orphans} we describe how we follow subhaloes after they can no longer be identified by our simulations (also known as ``orphan suhaloes''), and in section~\ref{sec:disruption} we explain our implementation to account for tidal disruption and stripping of stars.

\subsection{Orphan Implementation}
\label{sec:orphans}

Due to finite mass resolution of dark matter simulations, subhaloes can no be longer identified after they have lost a given fraction of their mass due to tidal striping. To improve convergence among simulation with different resolution, it is necessary to include ``orphan'' subhaloes/galaxy. These are structures with a known progenitor which we expect to still exist in the halo (i.e. that should not have merged with the central structure or completely destroyed via tidal forces).

Orphan sub-structures are normally used in semi-analytical models of galaxy formation (a.k.a. SAMs) to reproduce galaxy clustering \citep{DeLucia:2007,Guo:2011}. While there are some SHAM prescriptions that have included orphans in the past (eg. \citealt{Moster:2013}), most do not include this feature, because it adds an additional level of complexity to the model and to the underlying dark matter simulation. While uncommon, \cite{Guo:2014} found that not using an orphan prescription in the SHAM will cause some differences in the galaxy clustering prediction, even when using high-resolution simulations.

In this work, we include a flexible amount of orphan galaxies by tracking the most bound particle of subhalos with no known descendant. We assume a subhalo to be merged with their central structure only when the time since accretion exceeds the dynamical friction timescale of the subhalo. To compute the dynamical friction time, we use a modified version of the expression given by \cite[][Eq.7.26]{BT:1987}

\begin{equation}
\rm t_{dyn} = \dfrac{1.17\ t_{merge}\ d_{host}^2\ V_{host} (M_{host}/10^{13}\ h^{-1}M_{\odot})^{1/2}}{G\ ln(M_{host}/M_{sub}+1)\ M_{sub}},
\end{equation}

\noindent where $\rm t_{merger}$ is a free parameter that effectively regulates the number of orphan galaxies; $\rm d_{host}$ is the distance of the subhalo to the centre of its host halo; $\rm v_{host}$ is the virial velocity of the host halo; $\rm M_{host}$ is the virial mass of the host halo; and $\rm M_{sub}$ is the subhalo mass. The only variation from the original equation (besides the free parameter) is the term $\rm (M_{host}/10^{13}\ h^{-1}M_{\odot})^{1/2}$, which we find helpful to improve our predictions for the number of satellite galaxies in high halo masses.

\subsection{Tidal disruption}
\label{sec:disruption}

One of the most basic assumptions in SHAM is that the relation between stellar mass and subhalo property used ($\vpeak$ in our case) is constant through time and the same for central and satellites. While the $\vpeak$ - stellar mass relation is indeed very similar for centrals and satellites (c.f. Fig.~\ref{Fig:fig01}) it is not identical. In \cite{Smith:2016}, the authors showed that satellite galaxies can decrease their stellar mass to the point of disappearing into the intra-cluster medium. This effect should remove some satellite galaxies of a given sample (most of them located near the centre of a halo). 

To account for this effect, we follow \cite{Moster:2018} and assume that all the galaxies in subhaloes with current mass below a fraction $\Fs$ of their maximum subhalo mass during its evolution $\mpeak$ are disrupted,

\begin{equation}
\rm M_{sub} < f_s\ M_{peak}.
\end{equation}

In \cite{ChavesMontero:2016} the authors found that the clustering of a SHAM run in a dark matter simulation with the same resolution than EAGLE (a 67.7 $h^{-1}{\rm Mpc}$ high-resolution hydrodynamic simulation) has a larger clustering compare to the galaxies of EAGLE when selected by stellar mass. Similar results were found for the TNG300 by \cite{C20b} when doing SHAM mocks using $\vpeak$. This is because $\vpeak$ never decrease, contrary to the stellar mass of galaxies that can lose stellar-mass via stripping effect. By including a disruption parameter, the SHAM can reproduce better the galaxy clustering even in high-resolution simulations (see appendix A. for more details).

\subsection{Parameter variations}

We now explore the impact that these recipes and free parameters have in the predicted redshift space clustering of SHAM galaxies. Fig.~\ref{Fig:fig4} shows the multipoles and the projected correlation function of galaxies at $z=0$ with a number density of 0.01 $\ihMpcC$ in our TNG-300-mimic simulation. Coloured lines in each row show SHAM predictions varying either $\rm \sigma_{logM}$, the scatter between stellar mass and $\vpeak$; the merger timescale, $\rm t_{merger}$; or tidal disruption, $\rm f_{s}$; while keeping all other parameters fixed. For comparison, we also include the results for TNG galaxies as black lines.  

Firstly, we see that the $\rm \sigma_{logM}$ parameter has a small but systematic effect in the clustering. As its value increases, smaller subhaloes will be included in the selected sample which will progressively reduce the mean halo mass and large-scale bias. Note that the scatter we empirically measure from Fig.~\ref{Fig:fig01} is $\rm \sigma_{logM} \sim 0.125$.

Next we consider $\rm t_{merger}$ which has a significantly stronger impact. Large values of $\rm t_{merger}$ essentially imply that every subhalo ever accreted will still be present in the simulation -- this is thus an estimate for a simulation with infinite mass resolution. Consequently, there will be a large population of satellite galaxies in high-mass haloes, increasing its contribution and the clustering amplitude. In contrast, lower values of  $\rm t_{merger}$ implies orphans subhalos will be destroyed immediately, thus lowering the clustering and approaching the case of no orphan subhaloes. Note that this parameter brackets the two limiting cases of no artificial disruption in $N-$body simulations and of every subhalo destroyed being due to numerical reasons. From all the parameters introduced in this model, $\rm t_{merger}$ is the only one that can increase the galaxy clustering at small scales and is key to reproduce the galaxy clustering on low-resolution simulations.

Finally, as expected, larger (smaller) values of the tidal parameter $\rm f_s$ means fewer (more) satellite galaxies and thus, a lower (higher) clustering. Whereas this effect appears negligible compared to the effect of orphan subhalos and $\rm t_{merger}$, this parameter becomes more relevant when dealing with high-resolution simulations.  

We remind the reader all the changes show before are for a fixed number density, meaning that the effect on the galaxy clustering is not only due to the galaxies we are directly removing/adding, but also the ones that are enter/exit the sample to keep the total number of galaxy constant.

\section{Assembly bias}
\label{sec:ab}

The concept of halo assembly bias (also known as secondary bias, \citealt{Mao:2018}) was introduced by \cite{Sheth:2004} and \cite{Gao:2005} and is defined as the difference in the clustering of haloes of the same mass but of different secondary property (e.g. halo age, concentration, spin). Halo assembly bias can also propagate to galaxy clustering when haloes of the same mass but different secondary properties have a different halo occupation (effect know as occupancy variation, \citealt{Zehavi:2018,Artale:2018}). This difference in the galaxy clustering is known as galaxy assembly bias \citep{Croton:2007} and can cause differences in the correlation function of up to a 20\% for stellar mass-selected samples, depending on the redshift and number density of the galaxy sample in SAMs and hydrodynamic simulations \citep{C19,C20b}. 

In \cite{ChavesMontero:2016}, the authors found that the galaxy assembly bias signal predicted by SHAM can be up to 50\% lower than that measured in EAGLE. In \cite{C20b} we extended this work and found that hydrodynamic simulations, semi-analytical models, and SHAM predict different levels of galaxy assembly bias and that its evolution with redshift and number density is also different. Since it is unclear that the level of assembly bias in SHAM is a robust prediction, there we include an extension so that any amount of galaxy assembly bias signal can be added.

While the standard SHAM assumes that the scatter between the stellar mass and $\vpeak$ (or any other subhalo property used) is completely random (i.e. it does not depend on any secondary property of the subhalo), our method adds a correlation with the individual large-scale bias of subhaloes (measured with the method developed by \citealt{Paranjape:2018}). This is done for central and satellite galaxies independently, forcing conservation of the satellite fraction at all number densities (see section 4 of \citealt{C20b} for more details). 

To assign the level of galaxy assembly bias, the method uses two free parameters (one for central galaxies, $\rm A_c$, and one for satellite galaxies, $\rm A_s$) that control the correlation between stellar mass and individual bias of each subhalo at a given constant $\vpeak$ (one for central galaxies, $\rm A_c$, and one for satellite galaxies, $\rm A_s$). 

These free parameters can have a value between 1 (perfect correlation between stellar mass and bias-per-object, which will mean a maximum assembly bias signal), and -1 (perfect anticorrelation between stellar mass and bias-per-object, which will mean a minimum assembly bias signal), with 0 being a uncorrelated signal, meaning the sample will have the same assembly bias signal as the standard SHAM. Although not shown here, we kindly refer to Fig. 8 of \cite{C20b} where we showed that the difference in the galaxy clustering on large scales for a SHAM with maximum and minimum assembly bias is around a factor three. 


\section{Star formation rate}
\label{sec:sfr}

Most of the empirical models available to create mocks are focused on galaxy samples selected by stellar mass-like properties (e.g. HOD, SHAM, SCAM), and comparatively much less effort has been made to model star-forming galaxy samples. This is because most of these models relate a given (sub)halo property to a galaxy property, and while these relationships are quite simple for stellar-mass selected samples (i.e. the stellar mass is a good proxy for how massive a (sub)halo is) they are much more complicated for SFRs. More sophisticated models, such as semi-analytical models of galaxy formation, show no good agreement in how classic (sub) halo properties correlate with SFR-like galaxy properties \citep{C13,C15}.  

To account for this non-trivial relation between SFR-like galaxy properties and (sub) halo properties, empirical models have become more complicated when trying to reproduce these galaxies. Whereas a standard HOD \citep{Zheng:2005} can reproduce the galaxy clustering of a stellar-mass selected sample with reasonable accuracy using only $5$ free parameters, SFR-selected galaxy samples need HODs with 8 to 9 parameters for a reasonable agreement on the galaxy clustering \citep{GP:2020,Avila:2020}. Other empirical methods like the age matching or the conditional abundance matching (\citealt{Hearin:2013b,Hearin:2014, Kulier:2015}, Favole et al. in prep) have been able to reproduce the galaxy clustering using the additional correlation between secondary halo properties, such as halo age, and SFR, but with some limitations.

In this section, we implement a different approach to create star-forming mock galaxies. Inspired by
the fact that newly accreted mass provides gas to fuel the formation of new stars in galaxies and by the modelling carried out by {\tt EMERGE}, for central galaxies, we assume that the SFR is proportional to the dark matter mass accretion multiplied by an efficiency function. We assume that the star formation efficiency is maximal at a halo mass $\rm M_1$ (typically, $\sim10^{12}h^{-1}M_{\odot}$, close to the mass AGN feedback starts operating) and then it decays as a power law for both, more and less massive, host haloes. The SFR of the central galaxies can be expressed as:
\begin{equation}
\rm SRF_{centrals} \propto \dot{M}_{h} \dfrac{1}{(M_{h}/M_1)^\beta + (M_{h}/M_1)^{-\gamma}}
\end{equation}

For satellite galaxies, we assume the star formation rate is constant until they quench, with the same value they had when their subhalo mass was equal to its peak halo mass. For this, we stored the value of $\rm \dot{M}$ when $\rm M_{sh}=M_{peak}$ while running the $N$-body simulations. We consider a given satellite galaxy quenched when ${\rm t_{infall}}$, the time since accretion by a larger halo, is larger than a given quenching timescale:

\begin{equation}
{\rm t_{infall} < t_{quench} \equiv \tau_{0} \times  (M_{host\ halo}/10^{12}}h^{-1}{\rm M_{\odot})^{\tau_s}},
\end{equation}
with $\rm t_{dyn}$ the dynamical time of the halo ($\rm 0.1/H_{0}$), $\rm M_{host\ halo}$ the host halo mass of the subhalo and $\rm \tau_{0}$ \& $\tau_s$ free parameters. 

This last equation is slightly different from that used by {\tt EMERGE}. There, they assume a direct relation with the stellar mass of galaxies, while we assume a relation with their host halo mass. We find that, for our specific implementation, this describes better the SFR of TNG galaxies. Also, not using the stellar mass on the SFR calculation gives us more freedom when creating mocks with only SFR and not stellar mass.

To avoid including more free parameters to the mock, we will omit any normalisation parameter. This means that the value obtained for the SFR can not be compared with other values in the literature, but that a galaxy sample with a fixed number density should include the same galaxy population.

We show the impact of varying each of the SFR parameters in Fig.~\ref{Fig:fig4p5}. Similar to Fig.~\ref{Fig:fig4}, it shows the multipoles and the projected correlation function of galaxies at $z=0$ with a number density of 0.01 $\ihMpcC$ in our TNG-300-mimic simulation. Coloured lines in each row show the slopes in the star formation efficiency ($\beta$ \& $\gamma$), the halo mass of the peak of the star formation efficiency ($M_1$), the parameters that control de abundance of satellite galaxies $\tau_0$ and its dependence with it host halo mass $\tau_S$.

The dependence of the galaxy clustering of $\beta$ and $\gamma$ is weak, especially for $\beta$. Lower values of $\beta$ and higher values of $\gamma$ would mean selecting galaxies in more massive haloes. Since these haloes are more biased, the clustering of these galaxies should also be higher, as shown in the plot. The same thing happens with $M_1$, where a higher value of $M_1$ means more clustered galaxy sample, consistent in selecting more massive haloes. The dependence on the galaxy clustering with $\tau_0$ and $\tau_S$ is strong, with higher values of these parameters having a larger galaxy clustering. This makes sense since higher values of $\tau_0$ would mean more satellite galaxies, that have a strong effect on the clustering of galaxies, especially at small scales, and higher values of $\tau_S$ will put these galaxies in more massive haloes (that have more bias).

In the next section, we will show the results for the galaxy clustering for stellar mass and SFR mock galaxy clustering.

\section{Redshift-Space galaxy clustering predictions}
\label{sec:clustering}

\begin{figure*}
\includegraphics[width=\textwidth]{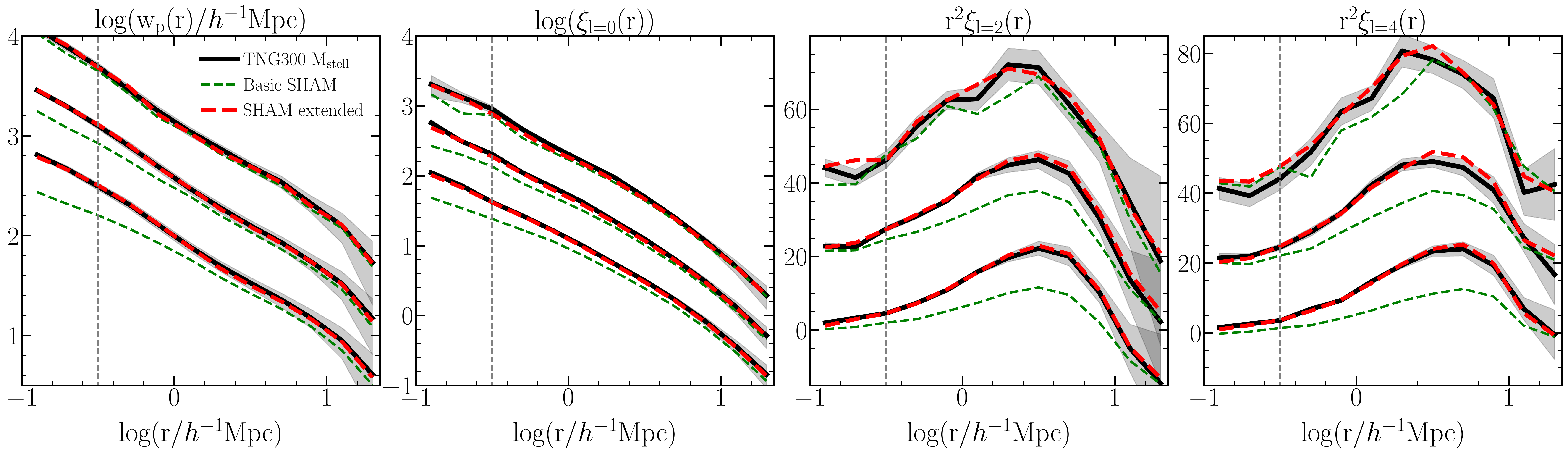}
\caption{
The projected correlation function ($\rm w_p(r_p)$, left most  panel), the monopole ($\rm \xi_{\ell=0}$, second panel), quadrupole ($\rm \xi_{\ell=2}$, panel row) and the hexadecapole ($\rm \xi_{\ell=4}$, right most) for  the TNG300 (black solid line), the basic SHAM (green dashed line) and or extended SHAM (red dashed line) for number density of n= 0.01, 0.00316 \& 0.001 $h^3{\rm Mpc^{-3}}$. Same as Fig.~\ref{Fig:fig02} the lowest number densities have been displaced to higher values for visualisation purposes, with the lowest number density at the top. The fit is done for scales above 0.3 $h^{-1}{\rm Mpc}$ (vertical dotted line).
}
\label{Fig:fig05}
\end{figure*}

In this section, we will compute the clustering predictions of our mocks and compare them to those in the TNG300. Since the simulation used to build the mocks has the same initial conditions and volume as the TNG300, cosmic variance should contribute little to the differences in clustering. 

To find the best set of values for the free parameters in our extended SHAM, we minimize the chi-squared ($\chi^2$) between the predicted projected correlation function ($w_p$), monopole ($\xi_{\ell=0}$), quadrupole ($\xi_{\ell=2}$), and hexadecapole ($\xi_{\ell=4}$) of the two-point correlation functions. We employ these statistics from 0.3 to 20 $\hMpc$. We compute uncertainties using 1000 jackknife samples, following the procedure of \cite{Zehavi:2002} \& \cite{Norberg:2009}. For simplicity, we assume no correlation between  different scales (i.e. we only use the elements of the diagonal of the covariance matrix). 

To minimise $\chi^2$, we use a particle swarm optimization (also known as PSO, \citealt{Kennedy:1995}). In this technique, a group of particles (known as the swarm) move with random velocities inside the  hyperspace we would like to explore. In each step, we compute the value of $\chi^2$ of the particular parameter combination. The velocities of the particles are then updated so they can point towards the location where each of the particle found its minimum $\chi^2$ and to the position with the minimum $\chi^2$ found by all the particle as a swarm. The amount of velocity that remains constant after each step, how strong is the deviation to the best local and global value, are the parameters that control how much ``exploration'' or ``exploitation'' the PSO will have. We chose a more ``explorative'' setup  using 16 particles with 400 steps each. The PSO typically converges in less than half of these steps.

\begin{table}
\begin{center}
 \begin{tabular}{c c c c c c} 
 \hline
 Stellar mass & TNG300 & & & & \\
 \hline
  ${\rm n}/{\ihMpcC}$ & $\sigmaLogM$ & $\tmerger$ & $\Fs$ & $\rm A_c$ & $\rm A_s$ \\ 
  \hline
  0.01 & 0.134 &  0.636 &  0.034 &  0.377 & -0.328 \\
  0.0316 & 0.193 & 0.406 & 0.003 & 0.227 &  -0.127\\
  0.001 & 0.059 & 0.765 & 0.081 & 0.328 & -0.032 \\
 \hline
  Stellar mass & SDSS & & & & \\
 \hline
    & $\sigmaLogM$ & $\tmerger$ & $\Fs$ & $\rm A_c$ & $\rm A_s$ \\ 
  \hline
    & 0.043 & 1.112 & 0.211 & 0.272 & 0.088 \\
 \hline
 SFR & TNG300 & & & & \\
 \hline
  ${\rm n}/{\ihMpcC}$ & $\beta$ & $\gamma$ & $\rm log(M_1)$ & $\rm \tau_0$ & $\rm \tau_s$ \\ 
  \hline
  0.01 & 2.490 & 5.127 & 12.770 &  4.931 & -0.363 \\
  0.0316 & 2.531 & 4.687 & 12.051, &   3.869 & -0.265\\
  0.001 & 8.166 &  8.907 & 12.063 & 4.346 & -0.307 \\
 \hline
  \end{tabular}
\end{center}
\caption{The best-fitting parameters of the extended SHAM for galaxies selected by stellar mass from the TNG300 (top part), the SDSS (middle part) and for galaxies selected by SFR of the TNG300 (bottom part). The fit is done for three different number densities (as labelled) for the TNG300 and all samples simultaneously for the SDSS.}
\label{Table:param}
\end{table}

\subsection{Stellar mass predictions}
\label{sec:clustering_mstell}

In Fig.~\ref{Fig:fig05} we show the clustering prediction of the TNG300 (black solid line) and our extended SHAM (red dashed lines) for three number densities, $n=0.01\ \ihMpcC$, $n=0.00316\ \ihMpcC$ \& $n=0.001\ \ihMpcC$ based on the stellar mass of the galaxies. The shaded black region represents 1 $\sigma$ uncertainty estimated from the jackknifes samples. For comparison, we show the clustering prediction of a standard SHAM with a scatter equal to that measured for TNG galaxies ($\sigma_{\log M}= 0.125$). 

The best-fitting parameters for our extended SHAM are shown in the upper section of Table~\ref{Table:param}. The two highest number densities have similar value for their parameters, with some small difference compared to the lowest number density. Note, however, that since these correspond to physically different galaxies, there is no a-priori a reason to expect identical values for the SHAM parameters.

We can see that our extended SHAM model can successfully reproduce the clustering of the TNG300 hydrodynamic simulation for all number densities. This is remarkable considering that the TNG300 was carried out with a full physics model and with a resolution 64 times higher than our DMO counterpart. We also notice that the extended SHAM has a better prediction compared to the standard SHAM, even for the highest number density. This is in agreement with previous results in the literature that predict that orphan galaxies are necessary for SHAM to reproduce the galaxy clustering, even in low number densities (e.g. \citealt{Guo:2014}). 

We further notice that the clustering is well reproduced at all scales, even below 0.3 $\hMpc$, the minimum scale used in the fitting of the SHAM parameters (indicated as a horizontal dotted line in the figure), suggesting the robustness of our model. We also notice that the clustering agrees better for the densest sample. Originally, we expected the opposite, since at these number densities the clustering depends more on the additional parameters we include (such as the fraction of orphans). We assume this is because at these scales we have a better statistic at larger number densities, making it easier to reproduce the general clustering of the sample.

\begin{figure*}
\includegraphics[width=0.35\textwidth]{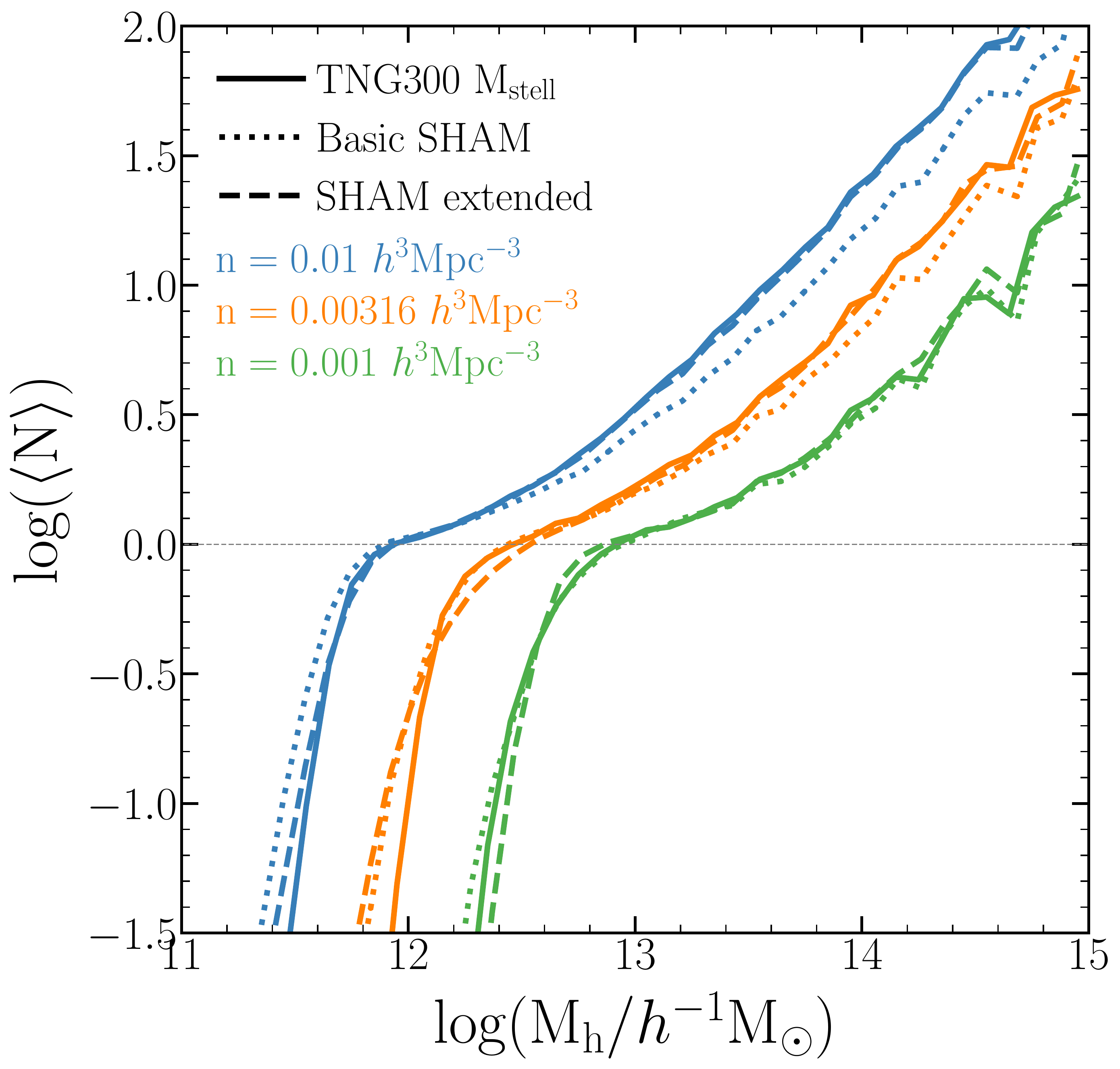}
\includegraphics[width=0.35\textwidth]{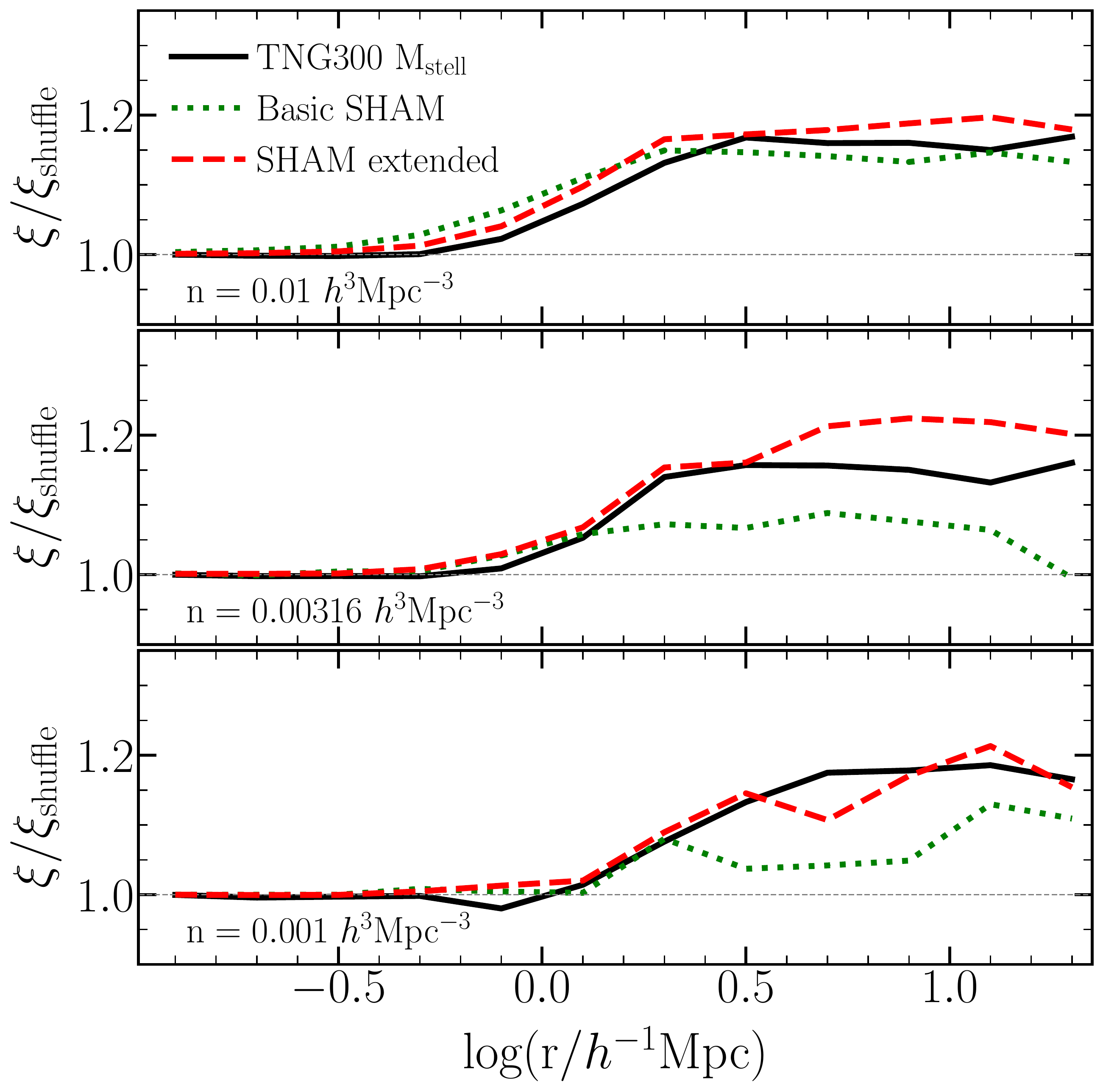}
\includegraphics[width=0.75\textwidth]{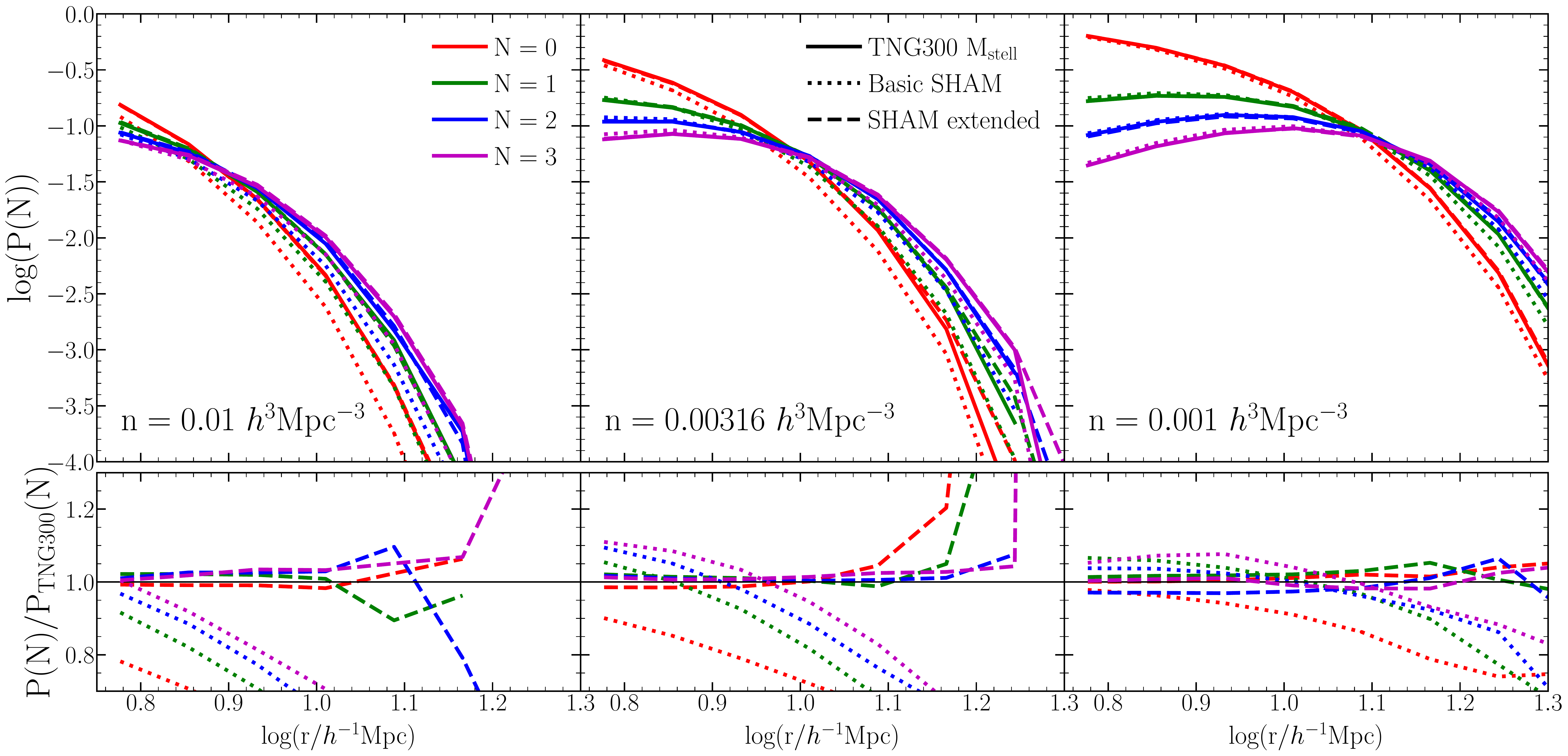} 
\caption{
{\it Top left:} The halo occupation distribution (HOD) for galaxies from the TNG300 (blue solid line), a basic SHAM technique (orange dotted line) and our extended SHAM model (green dashed line) for number densities of 0.01, 0.00316, and 0.001 $h^{3}{\rm Mpc^{-3}}$ based on their stellar mass, with the highest number density at the right of the plot. {\it Top right}: The galaxy assembly bias signal, measured as the ratio of the 2PCF of the selected galaxy sample to the one of its shuffled counterpart, following Croton et al. 2007. Same as the top panel, the solid black, dotted green and dashed red lines show the predictions for the TNG300, basic SHAM and our extended SHAM. {\it Bottom:} The upper part shows the k-nearest neighbour Cumulative Distribution Functions (kNN-CDF) for the TNG300 (solid lines) basic SHAM (dotted lines) and our extended SHAM model (dashed lines) for N = 0, 1, 2 \& 3 (red, green, blue and magenta line, respectively). The lower part of this row shows the ratio of the kNN-CDF of the standard and extended SHAM to the one of the TNG300.
}
\label{Fig:fig06}
\end{figure*}

To study the precision of the mock in more detail, we show in Fig.~\ref{Fig:fig06} the halo occupation distribution (HOD), the galaxy assembly bias (GAB), and the k-nearest neighbour Cumulative Distribution Functions (kNN-CDF) for the TNG300, our extended SHAM, and the standard SHAM for the same three galaxy samples as the previous figure. We emphasise that our model was not fitted to reproduce these statistics, and that its performance there is a direct reflection of its robustness.

We find that our extended SHAM model can successfully reproduce, in an almost perfect way, the mean number of galaxies ($\langle N \rangle$) that populate the haloes of mass ``$\rm M$'' in the TNG300, for all number densities. Small differences found at the transition of zero galaxies to one galaxy per halo are expected, since it is poorly constrained by galaxy clustering (see \citealt{Zehavi:2011}, on the constrain of $\sigma_{\log M}$, that control this transition\footnote{Notice that, while the $\sigma_{\log M}$ value of the HOD and of the SHAM are related, they are two different concepts.}). To properly compare the HOD from the TNG300 and the SHAMs from the TNG300-mimic, we match the halo masses of both simulations, following \cite{C15}.

The magnitude of GAB is estimated by computing the ratio between our galaxy sample, and that where the galaxy population was shuffled among haloes of similar masses, following the procedure of \cite{Croton:2007}. This method eliminates any dependence on the galaxy population and on any other property of the halo beyond its mass (i.e. any assembly bias component on the galaxy clustering). We remind the reader that, while two of our parameters regulate the amount of assembly bias of the galaxy population, we only fit these parameters the full galaxy clustering. The amount of assembly bias captured by the models is similar to the one of the TNG300, with sub-percental differences. This suggests that the assembly bias implementation in the SHAM done in \citep{C20b} can successfully retrieve the galaxy assembly bias of a galaxy sample. In a future work, we plan to use this model to constrain, among other things, the level of assembly bias from observational clustering measurements.

Finally, we test the k-nearest neighbour Cumulative Distribution Functions (kNN-CDF) of the three number densities for $k=0,\ 1,\ 2\ \&\ 3$. The kNN-CDF \citep{Banerjee:2020} measures the empirical cumulative distribution function of distances from a set of volume-filling Poisson distributed random points to the k–nearest data points. The performance of the extended SHAM shows to be highly accurate in all cases, with errors normally below the 5\% level. This is contrary to the case of the standard SHAM, where it never shows a good agreement. It is surprising that, even at the lowest number density (where it is normally assumed that the standard SHAM has a good performance and where the clustering and HOD shows a good agreement), the predictions of the kNN-CDF are poor. This shows us how important it is to introduce a proper abundance matching model even with high-resolution simulations and/or working with low number density samples.

\subsection{Star formation rate predictions}
\label{sec:SFR}

\begin{figure*}
\includegraphics[width=\textwidth]{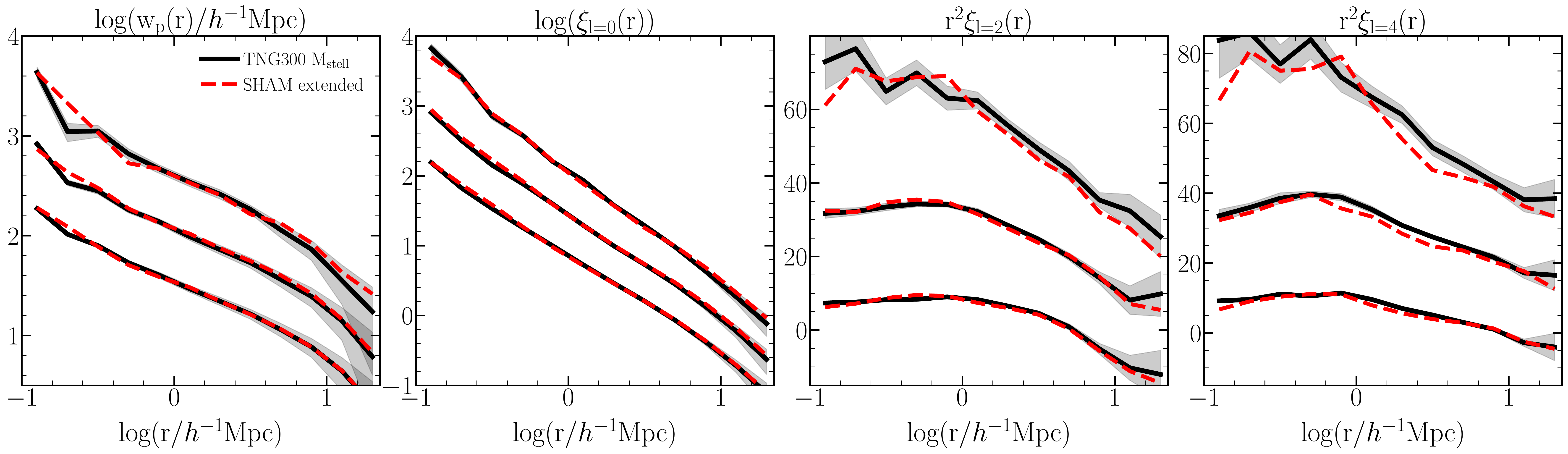}
\caption{
Similar to Fig.~\ref{Fig:fig05} but for galaxies selected by their SFR instead of stellar mass. We do not include predictions of the basic SHAM since this model does not predict SFR for it galaxies.
}
\label{Fig:fig07}
\end{figure*}

In Fig.~\ref{Fig:fig07} we show the clustering predictions for SFR-selected samples in our extended SHAM and in the TNG300. We do not include the basic SHAM in our results since it does not have any prediction for samples selected by SFR properties. While not as good compared to the stellar mass-selected sample, there is a good agreement with the TNG300, especially for the highest number densities. To be able to properly fit the clustering, we were on the need of using all scales above 0.1 $h^{-1} \mathrm{Mpc}$. The best-fitting parameters found for these number densities are shown in the lower part of Table~\ref{Table:param}. As for stellar mass, the two highest number densities have similar value for their parameters. The larger difference with the highest number density come from $\beta$ and $\gamma$, parameters we find have a low impact on the galaxy clustering, when fixing the number density of the sample.

\begin{figure*}
\includegraphics[width=0.35\textwidth]{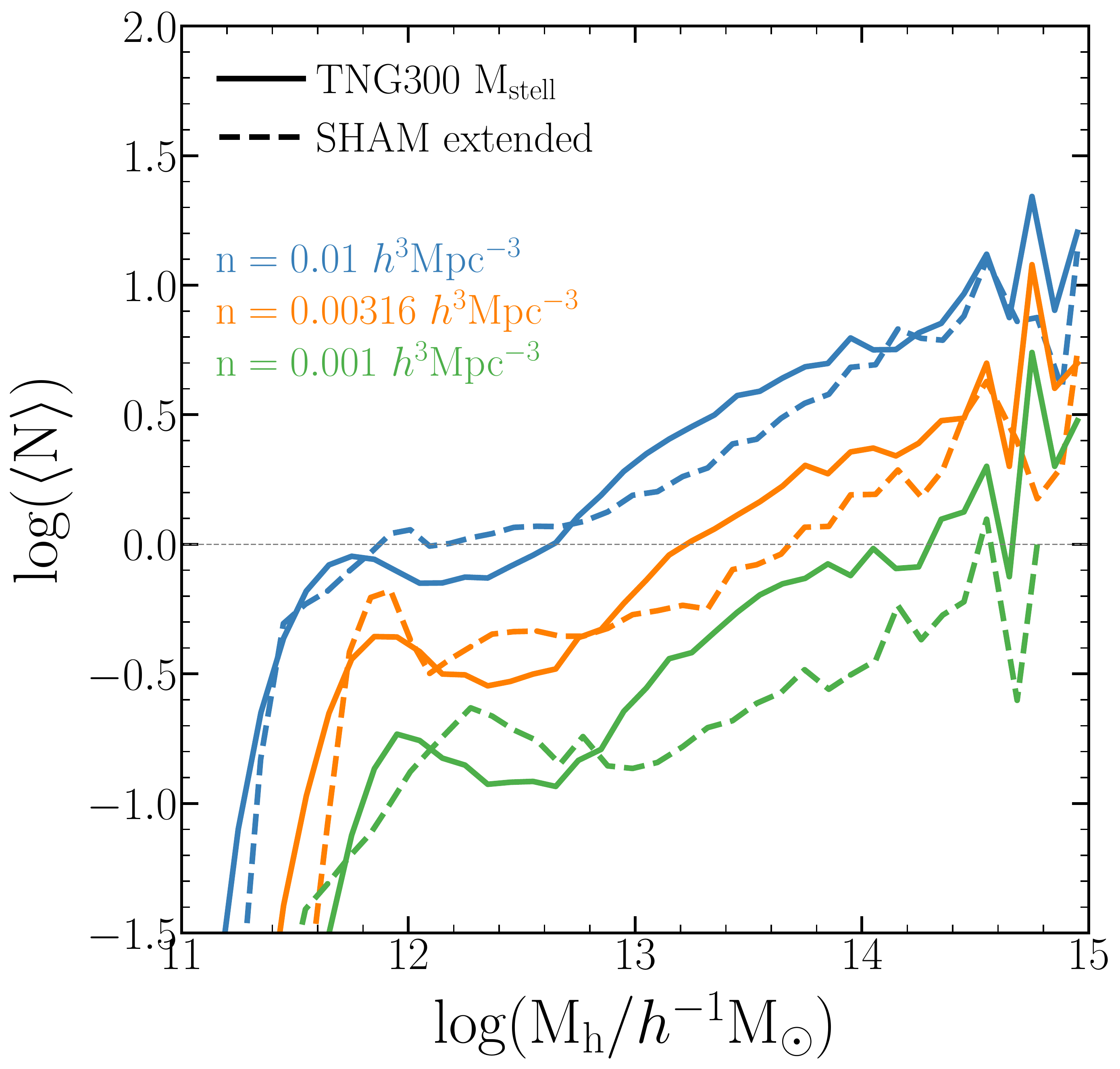}
\includegraphics[width=0.35\textwidth]{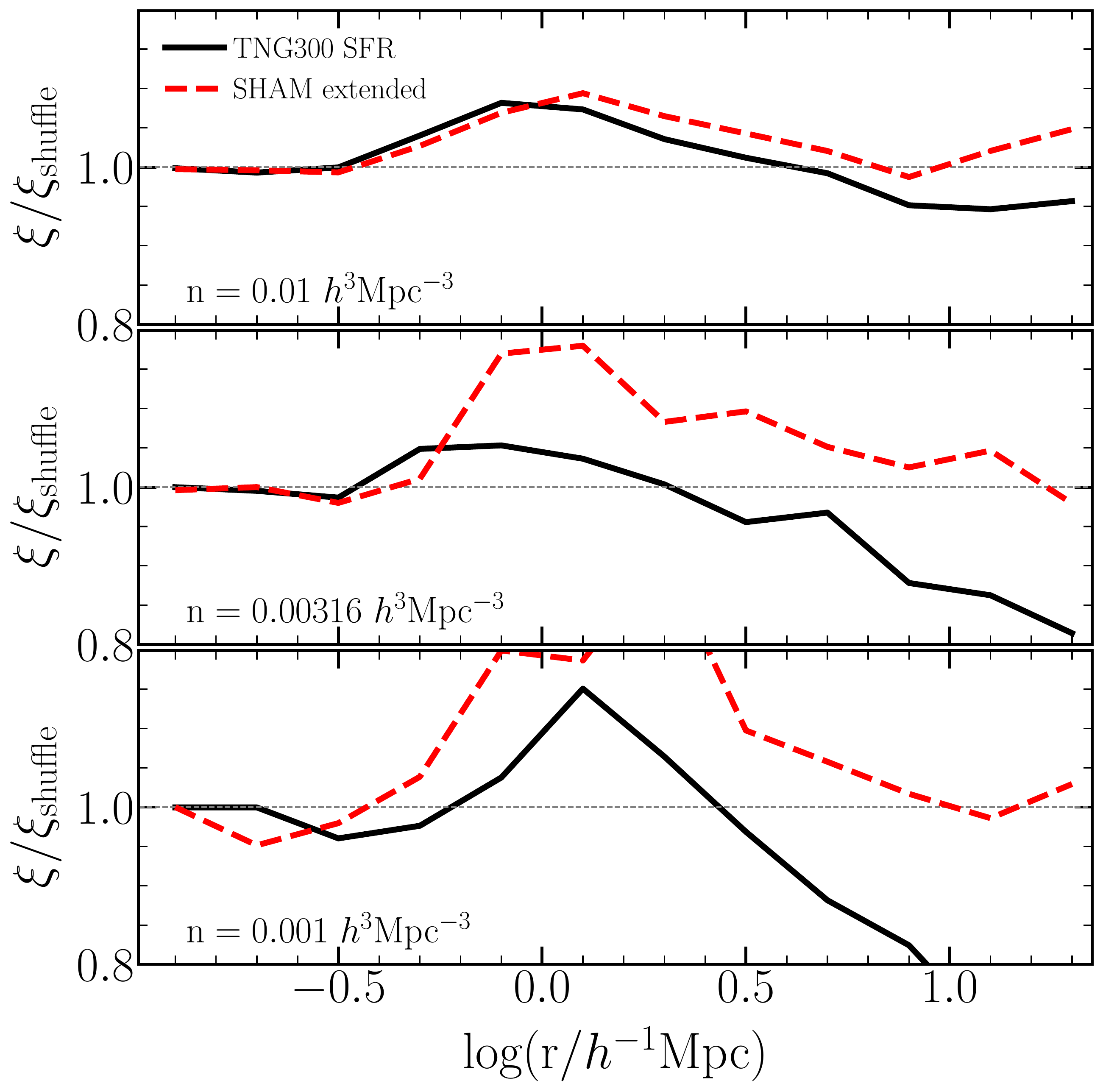}
\includegraphics[width=0.75\textwidth]{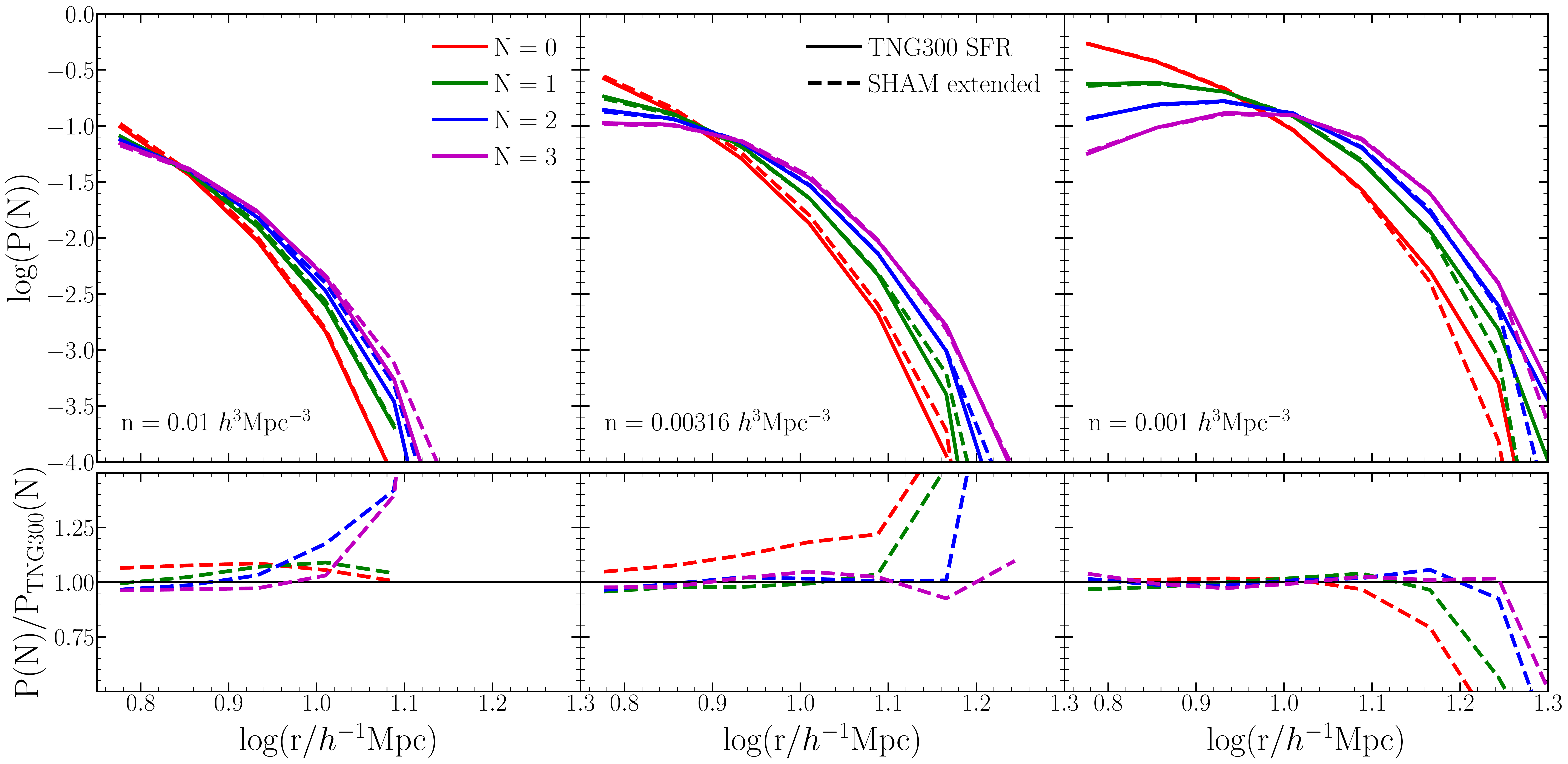}

\caption{
Similar to Fig.~\ref{Fig:fig06} but for galaxies selected by SFR instead of stellar mass, and without the predictions of the basic SHAM.
}
\label{Fig:fig08}
\end{figure*}

A further analysis of the galaxy samples is shown in Fig.~\ref{Fig:fig08}. As for the stellar mass-selected sample, we show the resulting HOD, GAB, and kNN-CDF. The HOD of the TNG300 and our extended SHAM show a similar shape, with some differences at high halo masses. These differences are small compared to those found by \cite{Avila:2020} when trying to fit HODs for a fixed bias, and to those found by \cite{C13} when comparing the HOD in different semi-analytical models. For galaxy assembly bias, the extended SHAM does not agree well with the TNG300. This is also expected since empirical models and different galaxy formation models do not share the same amount or even evolution of their galaxy assembly bias signal \citep{ChavesMontero:2016,C20b}. While for the stellar mass sample we added two additional parameters to explicitly model the level of assembly bias, we have not done so for SFRs. Adding more free parameters to an already complicated model could not be ideal, especially since the differences in the galaxy clustering between the extended SHAM model and the TNG300 are larger than the differences coming from the different galaxy assembly bias level. Finally, the kNN-CDF predictions show good agreement with the hydrodynamic simulation.

In summary, our SFR mocks show similar clustering properties as TNG300 galaxies. This agreement is not as good as for the stellar mass-selected sample, since the physics that dominates the star formation is much more complicated than those that rule the total stellar mass of galaxies. Nevertheless, the difference between our extended SHAM model and the TNG300 are small compared to the differences in  clustering among different galaxy formation models \citep{C13}.

\section{Prediction of the observed correlation function}
\label{sec:obs}

In the previous sections, we showed that our extended SHAM model is capable of reproducing the clustering of SFR- and stellar mass-selected galaxies of the TNG300 hydrodynamic simulation with high precision. Now, we aim to test the ability of our model to describe observed clustering data. 

We use the projected correlation function of galaxies in various stellar mass bins measured in the SDSS DR7 survey, as provided by \cite{Guo:2011} based on the work of \cite{Li:2006}. We fit our extended SHAM model for three ranges of stellar mass with ${\rm log(M_{stell}/}h^{-2}{\rm M_{\odot})}$ between [9.5 - 10], [10 - 10.5], \& [10.5 - 11]. Please notice that unlike  the rest of the paper, the mass unit is $h^{-2}{\rm M_{\odot}}$. To facilitate the comparison with TNG300 as well, we use its stellar mass function to build the SHAM.

\begin{figure*}
\includegraphics[width=0.95\textwidth]{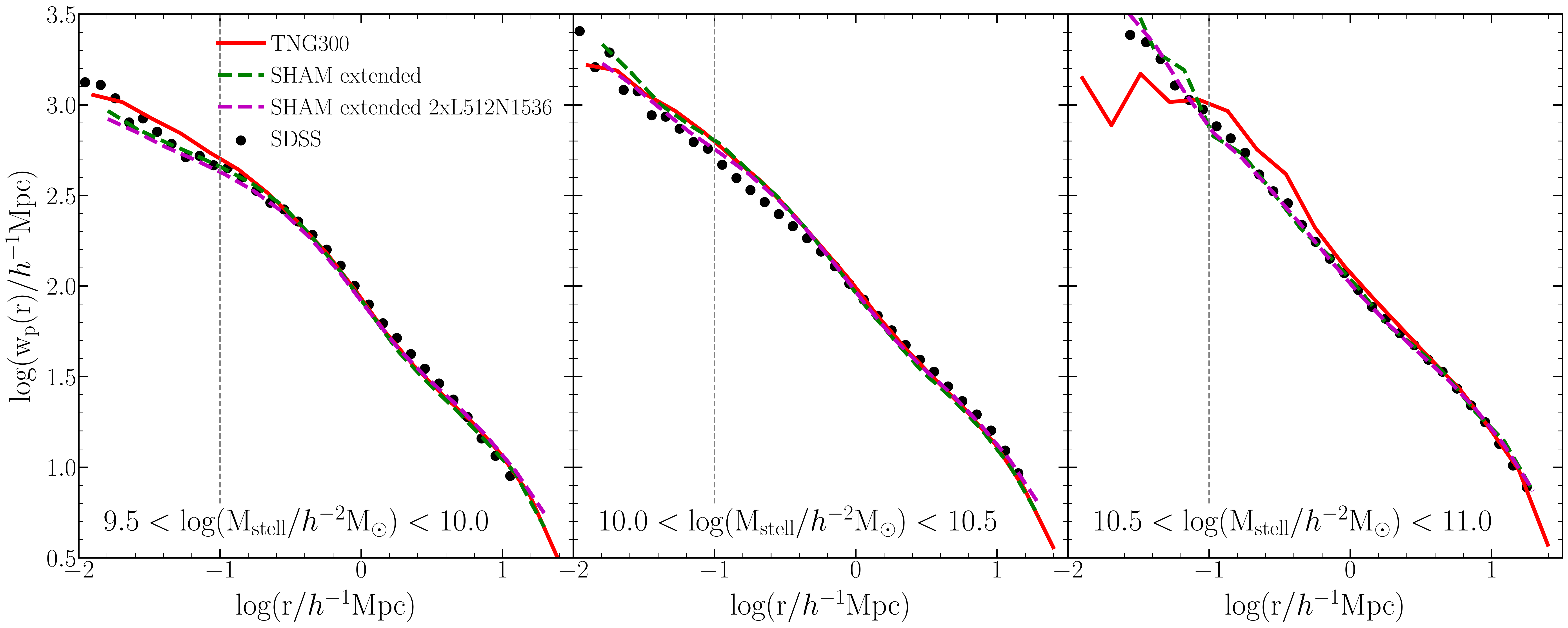}
\caption{
The projected correlation function, $w_p(r_p)$, for the TNG300 hydrodynamic simulation (red solid line), our extended SHAM implementation run over the TNG300-3 mimic (green dashed line) on a pair of simulations of 512 $\hMsun$ (using the average over the two resulting clustering measurements) and the measurements in the SDSS \citep{Guo:2011}. The right, middle and left panel show the predictions for galaxies with ${\rm log(M_{stell}/}h^{-2}{\rm M_{\odot})}$ between 9.5 - 10, 10 - 10.5 \& 10.5 - 11. Please notice that different from the rest of the paper, the units of mass are in $h^{-2}{\rm M_{\odot}}$. The SHAM was fitted so it can reproduce the SDSS prediction of the three different number densities at the same time}
\label{Fig:fig09}
\end{figure*}

The resulting projected correlation function is shown as a green dashed line in Fig.~\ref{Fig:fig09}. The observational data used to fit the extended SHAM model is shown as black circles, whereas the TNG300 is shown for comparison as a solid red line. For the fitting, we assume a constant arbitrary error per point, and use all scales above 0.1 $\hMpc$, the smallest scales at which we tested the extended SHAM model, shown as a vertical dotted line in the figure. All three mass range were fitted simultaneously (i.e. the predictions come from a single set of extended SHAM parameters). 

We see that our model shows good agreement with the observational data up to scales 10 times lower than the minimum scales used in the fitting, proving the flexibility and robustness of our model. The best-fitting parameters are shown in the middle part of Table~\ref{Table:param}. Notice that the two assembly bias parameters are positive, meaning that we required a larger amount of assembly bias signal compared to the standard SHAM to fit the observations. While this might point out to the existence of assembly bias in the Universe, a more detailed study is needed to properly address this.

Some of the small inconsistencies between our model and the SDSS are found at scales where the TNG300 and the SDSS are inconsistent too, with our model normally agreeing with the TNG300. Since our SHAM and the TNG300 trace the same cosmic structures (we recall both underlying simulations share the same initial white noise field), a possibility could be that the discrepancies originate from cosmic vartiance. To test for this, we apply our extended SHAM model to simulations with a significantly larger volume compared to the TNG300 ($512\ h^{-1}{\rm Mpc}$) but with a similar resolution ($1536^3$ particles) and cosmology (\citealt{Planck:2018}, without massive neutrinos). These simulations were run with opposite initial Fourier phases, using the procedure of \cite{Angulo:2016} that suppresses cosmic variance by up to 50 times compared to a random simulation of the same volume. 

The SHAM parameters were found using a PSO in our larger simulation, not the TNG300-mimic. The clustering prediction of the mean of the projected correlation functions, shown as  magenta dashed lines in Fig.~\ref{Fig:fig09}, however, are very similar to those obtained in the TNG300-mimic. This implies that cosmic variance in our simulations cannot explain the disagreement between SDSS observations and the theoretical predictions. The most logical explanation for these differences is that the different stellar mass functions between the TNG300 (that is the same as the SHAMs models) and the observational stellar mass function cause a different galaxy selection, or that it could be some measurement of the observational correlation function. Since we have shown the flexibility of this model by reproducing the observed clustering as a proof of concept, which was the goal of this test, we defer further analysis to future publications. 

\section{Summary and Conclusions}
\label{sec:Conclusions}

In this work, we have developed an extension of the standard subhalo abundance matchine technique (SHAM) that allowed us to reproduce the galaxy clustering in real and redshift space using relatively low-resolution simulations. 

The main characteristics of this model for stellar-mass selected samples are the following:

\begin{itemize}
 \item We start from a standard SHAM, that links the stellar mass of a galaxy sample with the peak circular velocity of the subhalo ($\vpeak$) in a 1-to-1 relation, and then add a scatter to it.

 \item We include an orphan model to the subhaloes to identify substructure even after they have fallen below a given mass resolution of a simulation. The amount of orphans is added with one free parameter, in a physically motivated model. 

 \item  Resolved subhaloes, that had lost a large fraction of their subhalo mass, are consider disrupted. 

 \item The scatter between the stellar mass and $\vpeak$ is modified to preferentially include subhaloes with a larger/lower large-scale bias (following the procedure of \citealt{C20b}). This changes the amplitude of the galaxy assembly bias signal of the sample.

\end{itemize}

For SFR-selected samples, we adopt the following model:

\begin{itemize}
 \item For central galaxies, and for a fixed value of halo mass, $\rm M_{200c}$, we assume the SFR is proportional to the accreted mass of the halo over the last snapshot ($\rm \dot{M_h}$). The dependence between the SFR and the halo mass (described in Eq. 3) is modeled as a double power-law with a maximum value at $M_1 = \rm M_{200c}$.

 \item For the satellite galaxies, we assume their SFR has been constant since their halo reached its peak mass ($\rm M_{200c} = M_{peak}$) until the galaxy quenched. The quenching is determined by as a function of the time elapsed since the galaxy became a satellite and of the host halo mass of the galaxy.
\end{itemize}

To test the performance of our model, we fitted the galaxy clustering of stellar mass and the SFR selected galaxies from the TNG300 magneto-hydrodynamic simulation at three different number densities: $n=0.01\ h^{3}\mathrm{Mpc}^{-3}$, $n=0.00316\ h^{3}\mathrm{Mpc}^{-3}$ \& $n=0.001\ h^{3}\mathrm{Mpc}^{-3}$, for the projected correlation function ($w_p$) and the monopole ($\xi_{\ell=0}$), quadrupole ($\xi_{\ell=2}$) and hexadecapole ($\xi_{\ell=4}$) of the redshift space correlation function (Fig.~\ref{Fig:fig05} and~\ref{Fig:fig07}). Our model successfully reproduces the galaxy clustering for all number densities considered, for both stellar mass and SFR selection (with higher precision for the former), and at $z=0$ adn $z=1$. This was done using a dark matter-only simulation with a resolution 64 times lower than the TNG300.

To further validate our mocks, we compared the halo occupation distribution (HOD), the amount of galaxy assembly bias (GAB), and the k-nearest neighbour Cumulative Distribution Functions (kNN-CDF) of our samples with those in the TNG300 (Fig.~\ref{Fig:fig06} \&~\ref{Fig:fig08}). We emphasise that none of these statistics were used in fitting the free parameters of our extended SHAM model. For the stellar mass sample, all the mock predictions agree almost perfectly with those of TNG300 galaxies. 
For the SFR selection, the kNN-CDF and the HOD show a reasonably good agreement with TNG300 galaxies. The discrepancies found here and in the galaxy clustering are small compared to the differences between different state-of-the-art galaxy formation models (eg. \citealt{C13}) and is mainly because the SFR is a complex and uncertain quantity to model, and thus, difficult to parametrise with some basic subhalo information. The amplitude of the galaxy assembly bias of the SFR-selected samples do not agree particularly well with those measured in the TNG300, but the overall differences are much smaller than those seen in the galaxy clustering. To keep the model simpler, we decided not to add more parameters and attempt to improve the agreement.  

Finally, to show the flexibility of our model, we reproduced the projected correlation function of a galaxy sample of the SDSS selected by stellar mass. We again found a good agreement with their clustering measurement (Fig.~\ref{Fig:fig09}), with most of the differences coming probably for the particular stellar mass function used by our model (that is the stellar mass sample of the TNG300).

We conclude that this {\bf S}ubhalo {\bf A}bundance {\bf M}atching {\bf e}xtended model (SHAMe) has the flexibility to reproduce with a high accuracy most stellar mass or SFR selected samples, independently if they are based on models or observations. On a follow-up work, we plan to use our SHAMe model to populate galaxies in scaled simulations \citep{Angulo:2020, Zennaro:2019}, following the procedure of \cite{C20a}, to constrain cosmological information from current and future galaxy surveys.

\section*{Data availability}
The data underlying this article will be shared on reasonable request to the corresponding author.

\section*{Acknowledges}

We thank useful comments from Giovanni Aric\`o, Jon\'{a}s Chaves Montero, and Marcos Pellejero.
The authors acknowledge the support of the ERC Starting Grant number 716151 (BACCO). SC acknowledges the support of the ``Juan de la Cierva Formaci\'on'' fellowship (FJCI-2017-33816). The authors also acknowledge the computer resources at MareNostrum and the technical support provided by Barcelona Supercomputing Center (RES-AECT-2019-2-0012 \& RES-AECT-2020-3-0014)".

\bibliography{Biblio}

\appendix

\section{Tidal disruption on high resolution simulations}

\begin{figure*}
\includegraphics[width=1.00\textwidth]{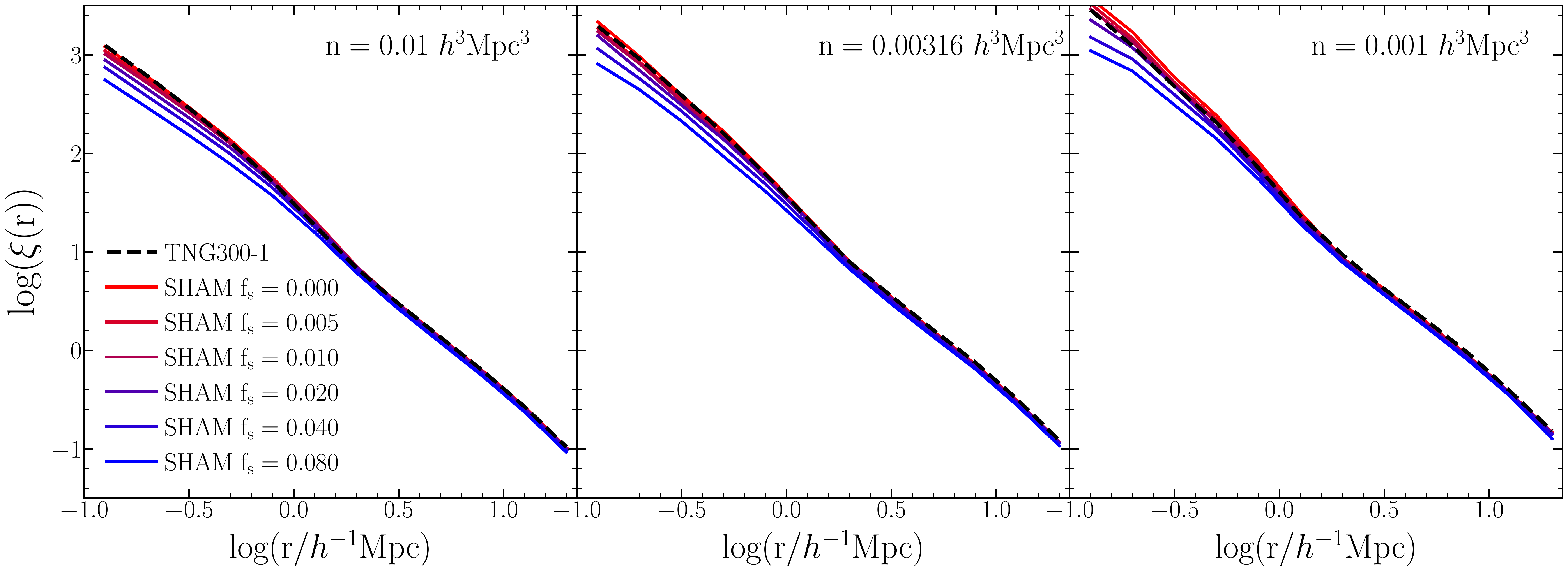}
\caption{
Similar to the right panel of Fig.~\ref{Fig:fig4} but only for the 2PCF ($\rm \xi(r)$) with the SHAMs runs over the TNG300-1-Dark simulation.
}
\label{Fig:figA1}
\end{figure*}

In section~\ref{sec:disruption} we show the impact of the stripping parameter $\rm f_s$ on the galaxy clustering. While the efect of this parameter on the galaxy clustering is lower compare to the rest of the parameters implemented here, its impact becomes more important on high-resolution simulations, where the galaxy clustering can be overestimated by the standard SHAM (lowest number density of Fig~\ref{Fig:fig02}). This same effect is shown in Fig.~\ref{Fig:figA1} where we show the correlation function of a standard SHAM (no orphans or additional galaxy assembly bias) with different values of $\rm f_s$ run over the TNG300-1-Dark only dark matter simulation and compare it with the TNG300 full hydrodynamic run. The scatter of these SHAM is $\sigma_{logM}=0.125$, similar to the one expected by the TNG300. We notice that for the most massive sample (lower number densities) a value of $\rm f_s=0$ (meaning no disruption), can not fit the clustering of the hydrodynamic simulation. This can only be achieved using larger values of $f_s$.

\section{Predictions at higher redshift}
\begin{figure*}
\includegraphics[width=1.0\textwidth]{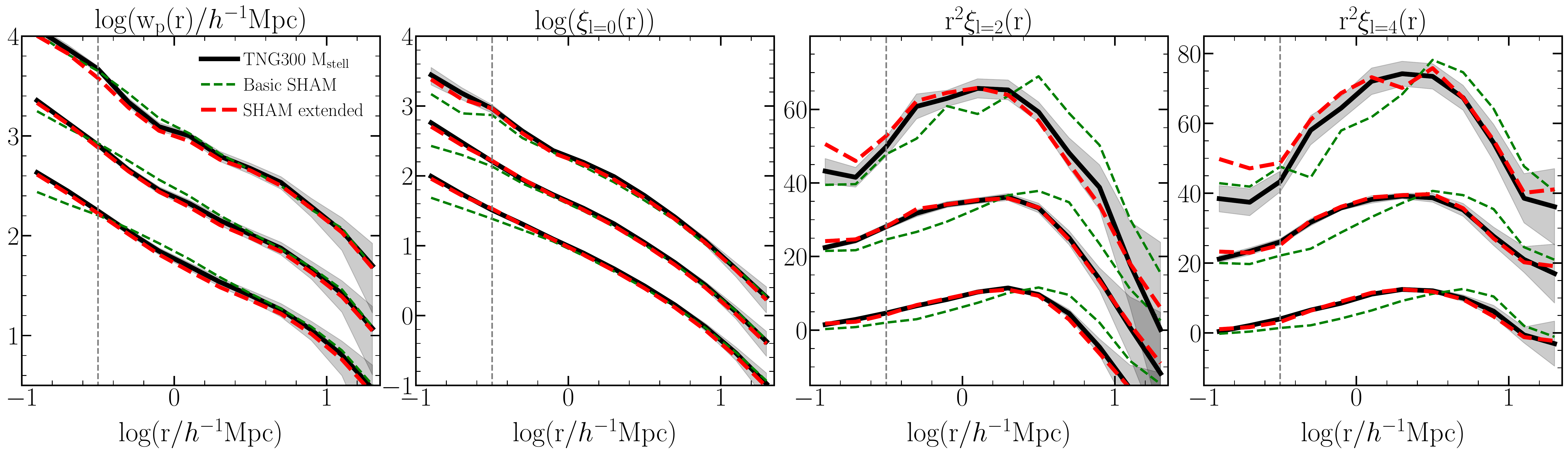}
\caption{
Similar to Fig.~\ref{Fig:fig05}, but for galaxies at z=1 instead of z=0.
}
\label{Fig:figA2}
\end{figure*}

\begin{figure*}
\includegraphics[width=1.0\textwidth]{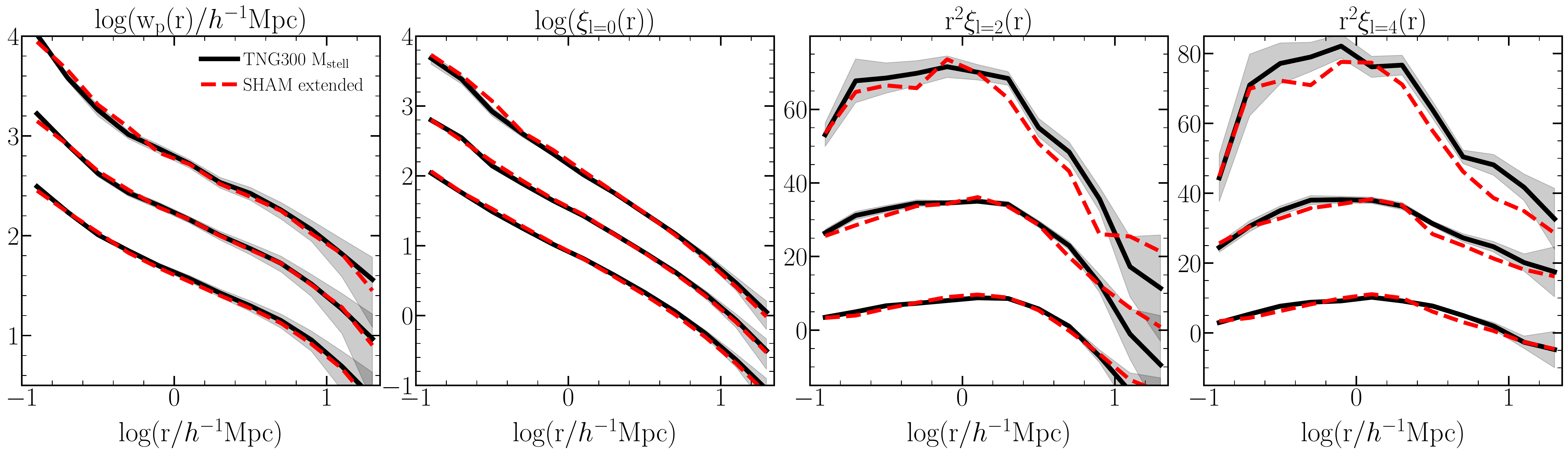}
\caption{
 to Fig.~\ref{Fig:fig07}, but for galaxies at z=1 instead of z=0.
}
\label{Fig:figA3}
\end{figure*}

In this paper, we showed the performance of our extended SHAM model by comparing its predictions to the TNG300 hydrodynamic simulation. While we only showed the predictions at $z=0$, we also test its performance at higher redshifts. On Fig.~\ref{Fig:figA2} and Fig.~\ref{Fig:figA3} the prediction of our extended SHAM model when fitting the clustering of the TNG300 at z=1 for stellar mass and SFR selected galaxies. The performance for the stellar mass-selected sample is similar to the one at z=0 (Fig.~\ref{Fig:fig05}) while the predictions for the SFR selected galaxies look slightly better than for z=0 (Fig.~\ref{Fig:fig07}), proving the flexibility and robustness of our model.

\bsp	
\label{lastpage}
\end{document}